\documentclass[aps, superscriptaddress, showpacs, floatfix, titlepage]{revtex4-1}

\usepackage[dvipdfmx]{graphicx}
\usepackage{epsfig}
\usepackage{graphicx}
\usepackage{amsmath}
\usepackage{footmisc}
\usepackage{hyperref}
\usepackage{multirow}
\usepackage{graphicx}
\usepackage{amsmath}
\usepackage{feynmf}
\usepackage{xcolor, cancel}
\usepackage[normalem]{ulem}  

\usepackage[sort,compress]{cleveref}
\crefname{section}{Sec.\!}{Secs.\!}
\crefname{equation}{Eq.\!}{Eqs.\!}
\crefname{figure}{Fig.\!}{Figs.\!}
\crefname{table}{Tab.\!}{Tabs.\!}
\crefname{appendix}{App.\!}{Apps.\!}

\newcommand{\Ex}[2]{\ifmmode{#1\times10^{#2}}\else{$#1\times10^{#2}$}\fi}

\begin{document}

\title{Functional Renormalization Group analysis of the quark-condensation pattern on the Fermi surface: A simple effective-model approach}

\author{Kie Sang Jeong}
\email{jeong@itp.uni-frankfurt.de}

\author{Fabrizio Murgana}
\email{fabrizio.murgana@dfa.unict.it}

\author{Ashutosh Dash}
\email{dash@itp.uni-frankfurt.de}

\author{Dirk H.\ Rischke}
\email{drischke@itp.uni-frankfurt.de}

\affiliation{Institute for Theoretical Physics, Goethe University, Max-von-Laue-Str.1, D-60438 Frankfurt am Main, Germany}

\date{\today}

\begin{abstract}

A simple effective model for the intermediate-density regime is constructed from the high-density effective theory of quantum chromodynamics (QCD). 
In the effective model, under a renormalization-group (RG) scaling towards low momenta, the original QCD interactions lead to four-quark contact interactions for the relevant quark and hole modes around the Fermi surface. 
The contact interaction in the scalar channel can be traced back to zero-sound-type collinear quark scattering near the Fermi surface in an instanton background. 
The quark and hole states in opposite directions of a given Fermi velocity form the collective scalar bosonic mode $\sigma$. 
The magnitude of $\sigma$ is investigated via the non-perturbative Functional Renormalization Group (FRG) evolution of the effective average action from the ultraviolet (UV) to the infrared (IR). 
In the mean background-field approximation for $\sigma$, nontrivial minima ($\bar{\sigma} \neq 0$) are found in the IR limit of the effective average action. 
A nonvanishing $\bar{\sigma}$ corresponds to condensation of  quark and hole states in opposite directions of a given Fermi velocity, in a thin shell-like structure in momentum space around the Fermi surface.
This looks similar to the shell-like baryon distribution in momentum space assumed in the quarkyonic-matter concept. 
However, when including a dynamic bosonic $\sigma$-mode in the RG flow, we find that its diffusive nature destroys the quark-hole condensate, i.e., the IR potential does not show any minima beyond the trivial one.

\end{abstract}

\maketitle 

\section{Introduction}

The scalar quark condensate $\left \langle \bar{\psi} \psi \right\rangle$ plays an important role in the descripton of the low-energy ground state of QCD since it is the order parameter related to spontaneous chiral symmetry breaking. 
The dynamically generated constituent quark mass in a hadronic state is determined by the chiral order parameter in low-energy effective models of QCD~\cite{Nambu:1961tp, Nambu:1961fr, Gell-Mann:1960mvl, Gross:1974jv}. 
Also, in QCD sum-rule analyses of hadronic states~\cite{Shifman:1978bx, Shifman:1978by}, the ground-state properties are mainly determined by the quark condensates appearing in the operator-product expansion of the correlation function of the hadron interpolating field.  
At low nuclear matter densities, the attractive scalar self-energy of the quasi-nucleon state leads to a reduced magnitude of the in-medium quark condensate~\cite{Drukarev:1988ib, Cohen:1994wm}. 
If one assumes the quasi-particle picture
and, correspondingly, a homogeneous condensation amplitude in the nuclear medium like in vacuum, such a decreasing tendency~\cite{Cohen:1991nk, Fiorilla:2012bc} implies that, above some density, the quark condensate vanishes and a phase transition to quark matter occurs.

On the other hand, it was speculated that an inhomogeneous configuration can emerge in dense nuclear matter, where translation invariance of the system is broken, rather than having a direct phase transition to quark matter. 
Density waves in dense nuclear matter were discussed by Overhauser~\cite{Overhauser:1960zz} and a series of studies about inhomogeneous pion condensation was reported in Refs.~\cite{Weise:1975tk, Migdal:1978az, Migdal:1973zm}. 
Recently, the quarkyonic-matter concept~\cite{McLerran:2007qj} emerged from large-$N_c$ QCD~\cite{tHooft:1973alw} and has been applied to dense nuclear matter~\cite{McLerran:2018hbz, Jeong:2019lhv, Zhao:2020dvu, Duarte:2020xsp, Duarte:2020kvi, Koch:2022act, Xia:2023omv, Poberezhnyuk:2023rct, Dey:2024lco}.
The shell structure of quarkyonic matter in momentum space implies a periodic distribution of baryon number in configuration space. 
If the quasi-particle distribution is inhomogeneous in configuration space, the corresponding quark-condensation pattern should have a non-trivial inhomogeneity as well. 
The possibility of inhomogeneous quark condensation was first discussed in the large-$N_c$ limit~\cite{Deryagin:1992rw}.
Subsequent studies investigated whether inhomogeneous quark condensation can  compete with the color-superconducting phase~\cite{Shuster:1999tn, Park:1999bz, Rapp:2000zd}.  
Discussions within effective models have been reported as well~\cite{Nakano:2004cd,Schnetz:2004vr,Nickel:2009ke, Nickel:2009wj, Kojo:2009ha, Basar:2009fg, Buballa:2014tba, Buballa:2018hux, Lenz:2020bxk, Lenz:2020cuv, Koenigstein:2021llr}.

\begin{figure}
\includegraphics[height=5.5cm]{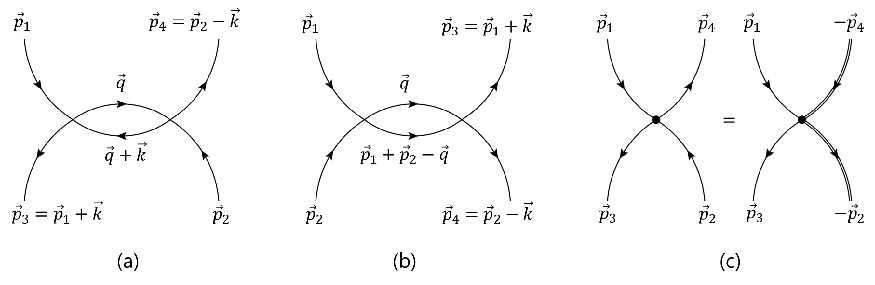} 
\caption{(a) Zero-sound-type diagram, (b) BCS-type diagram, and (c) effective four-quark interaction diagram, which gets contribution from both (a) and (b). 
In (a) and (b), the incoming momenta are $\vec{p}_1$ and $\vec{p}_2$, while the outgoing momenta are $\vec{p}_3$ and $\vec{p}_4$.
These diagrams become dominating when the fermion lines in the loop hit the singular point at the  Fermi surface simultaneously. For (a),  when the transferred momentum is minimal ($\vec{k}\simeq 0$), both internal lines become singular at $\vert \vec{q} \vert \rightarrow \vert \vec{p}_F \vert$.
When scaling to the Fermi surface, the collinear scattering ($\vec{p}_1 \simeq \vec{p}_3$, $\vec{p}_2 \simeq \vec{p}_4$, $\vec{p}_1 \simeq \pm \vec{p}_2$, $\vert \vec{p}_i \vert =\vert \vec{p}_F \vert$) becomes marginal in the Wilsonian sense.  For (b), both internal lines become singular at $\vert \vec{q} \vert \rightarrow \vert \vec{p}_F \vert$ when the incoming momenta are on opposite sides of the Fermi surface ($\vec{p}_1\simeq -\vec{p}_2$). 
For the effective description (c), if the incoming $\vec{p}_2$ line and the outgoing $ \vec{p}_4$ line are conjugated, then the process can be understood as the forward scattering of an incoming particle with momentum $\vec{p}_1$ and a hole with momentum $-\vec{p}_4$ to an outgoing particle with momentum $\vec{p}_3$ and a hole with momentum $-\vec{p}_2$ with minimal momentum transfer ($\vec{k}\simeq 0$). 
For Overhauser-type scalar quark pairing, the corresponding Fermi velocities in Eq.~\eqref{4qeff} should be $\vec{v}_1 \simeq -\vec{v}_3 \simeq -\vec{v}_2 \simeq \vec{v}_4$, since $\Gamma,\Gamma^{`}  = I$ allows only for such a configuration.}\label{f}
\end{figure}

In the cold, dense limit ($\mu_{q} \gg \Lambda_{\textrm{QCD}}$, $T\ll  \Lambda_{\textrm{QCD}}$), the (3+1)-dimensional QCD Lagrangian can be reduced to a longitudinal high-density effective theory (HDET)~\cite{Hong:1998tn, Hong:1999ru, Nardulli:2002ma, Schafer:2003jn}. 
At intermediate densities  ($\mu_{q} \gtrsim \Lambda_{\textrm{QCD}}$), where non-perturbative gauge-field interactions become dominant, one may construct an effective model analogous to the Gross-Neveu (GN) model~\cite{Gross:1974jv} in terms of the relevant quark and hole modes in the collinear direction appearing in HDET. 
In this effective model, QCD interactions lead to instanton-mediated four-quark interaction terms on the Fermi surface. 
In this work, we investigate the possible condensation pattern of quark modes around the Fermi surface via bosonization of the four-quark interaction. 
The general form of the four-quark interaction defined around the Fermi surface can be expressed as follows~\cite{Nardulli:2002ma, Schafer:2003jn}:
\begin{align} 
\mathcal{L}_{\textrm{4q}} & = \sum_{\vec{v}_{i}} \sum_{\Gamma, \Gamma'}\frac{c^{\Gamma \Gamma'}(\vec{v}_1,\vec{v}_2,\vec{v}_3,\vec{v}_4)}{\mu^2}\left[ \bar{\psi}_{+}(\vec{v}_3, x)\Gamma\psi_{+} (\vec{v}_1, x) \right] \left[\bar{\psi}_{+}(\vec{v}_4, x) \Gamma' \psi_{+}(\vec{v}_2, x) \right]+\ldots\;,\label{4qeff}
\end{align}
where $\Gamma$ and $\Gamma'$ denote elements of the Clifford basis for the quark bilinears. 
The fields $\psi_{+} (\vec{v}_i, x)$  denote the quark modes around the Fermi surface, where $\vec{v}_i$ is the Fermi velocity, whose magnitude is given as $\vert \vec{v}_{i} \vert = \vert \vec{p}_{Fi} \vert / \sqrt{\vec{p}_{Fi}^{\,2} + m_i^2} $. 
Two types of collinear scattering processes become marginal in the Wilsonian sense when scaling the momentum exchange towards the Fermi surface: the zero-sound channel, cf.~Fig.~\ref{f}(a), and the Bardeen-Cooper-Schrieffer channel, cf.~Fig.~\ref{f}(b), both of which satisfy the collinear scattering conditions~\cite{Nardulli:2002ma}.
In this work, we consider the Overhauser-type scalar pairing, for which $\vec{v}_1 \simeq -\vec{v}_3 \simeq -\vec{v}_2 \simeq \vec{v}_4$.
This channel gets contributions from both the zero-sound and BCS-type diagrams, and the zero-sound type vector pairing. 
The BCS-type channel leading to color superconductivity will be left for future work.

Since we define the effective model with non-perturbative four-quark interactions at intermediate densities, the effective average action of the system will be investigated in the Functional Renormalization Group (FRG)  framework~\cite{Wetterich:1991be, Wetterich:1992yh, Ellwanger:1993mw, Morris:1993qb, Morris:1994ki, Berges:2000ew, Dupuis:2020fhh}, instead of a renormalization-group (RG) analysis of the couplings at a given finite order of the perturbative expansion. 
In the FRG analysis, quantum and thermal fluctuations are integrated out successively from an energy or momentum scale in the UV to the IR. 
The RG flow of the effective average action corresponds to a  successive functional integration over the quantum fields.
It can be formulated via an exact flow equation for the effective average action at a given scale $k$.
The flow equation describes the evolution of the effective average action in ``RG time'' $t \equiv -\ln{ k/\Lambda_{\textrm{UV}}}$, where $\Lambda_{\textrm{UV}}$ is the UV scale. 
In this approach, the effective four-quark interactions can be expressed in terms of  bosonic collective fields via a Hubbard-Stratonovich transformation~\cite{1957SPhD....2..416S, Hubbard:1959ub}, where the bosonic fields have the quantum numbers of the mesons, which are the dominant degrees freedom in the IR limit. 
The bosonized Lagrangian corresponds to the UV limit of the effective average action and its RG flow to the IR limit can be obtained in both the mean background-field approximation and for a dynamical scalar boson field $\sigma$ within the FRG framework. 
Considering a dynamical $\sigma$-mode, the derivative of the exact flow equation with respect to $\sigma$ is a partial differential equation of advection-diffusion type known from fluid dynamics~\cite{Grossi:2021ksl, Stoll:2021ori, Koenigstein:2021rxj, Steil:2021cbu,Murgana:2023xrq,Murgana:2023pyx}. 
Thus, methods well-known from fluid dynamics can be utilized to solve the flow equation. 
If the  IR potential develops  a non-trivial minimum $\bar{\sigma} \neq 0$ in a momentum shell around the Fermi surface, quark and hole states with opposite momenta on the Fermi surface form a scalar condensate, which breaks the chiral symmetry among the relevant quark modes.
The broken phase in the momentum shell around the Fermi surface is related to an exotic baryon-number distribution in the chiral density-wave picture~\cite{Overhauser:1960zz, Deryagin:1992rw} and the quarkyonic-matter concept~\cite{McLerran:2007qj}.

This paper is organized as follows: In Sec.~\ref{sec2}, the definition of quark and hole states appearing in HDET will be introduced. 
The possible condensation patterns in an instanton background and the QCD origin of the contact interactions in the effective model will be discussed as well. 
In Sec.~\ref{sec3}, the RG flow of the average effective action of the  model will be analyzed in the FRG framework: first using the mean-field approximation for the $\sigma$-mode, including both the homogeneous and the inhomogeneous quark-hole condensation, and later solving the FRG flow equation for a dynamical $\sigma$-mode~\cite{Stoll:2021ori, Koenigstein:2021rxj, Steil:2021cbu,Murgana:2023xrq,Murgana:2023pyx}.
Concluding remarks are given in Sec.~\ref{sec4}.

\section{High-density effective Lagrangian and Fermi-surface modes}~\label{sec2}

In this section, we introduce the quark and hole modes which are excited in the direction longitudinal to the Fermi momentum appearing in HDET.
The possible condensation patterns between the relevant surface quark modes, which can be considered explicitly in an effective approach analogous to the GN model, will be introduced briefly. 
The possibility of surface zero-mode correlations in an instanton background will be discussed as well, since the contact four-quark interaction appearing in a simple model (which will be introduced in Sec.~\ref{sec3}) can be traced back to the four-quark interaction between the surface quark modes in an instanton background.

\subsection{Positive- and negative-energy mode decomposition}

Our starting point is isospin-symmetric, high-density QCD matter, at a quark chemical potential $\mu \gg m_q, \Lambda_{\textrm{QCD}}$, such that we can take quarks to be massless and weakly interacting.
The Lagrangian of this system is
\begin{align}
\mathcal{L}_{\textrm{QCD}} & =  -\frac{1}{4}\mathcal{F}^{a}_{~\mu\nu}\mathcal{F}^{a\mu\nu}+ \bar{\psi} (i \,\slash \hspace{-0.25cm}  D +
\mu \gamma^0) \psi\;,\label{qcdl}
\end{align}
where $\mathcal{F}_{\mu\nu}=(-i/g)[D_{\mu}, D_{\nu}]$, with $D_{\mu}=\partial_{\mu} +ig A_{\mu}$, and a summation over the $N_f$ quark flavors is implied.
At low temperatures, low-energy quark and hole states around the Fermi surface are excited mainly in the direction of the Fermi velocity $\vec{v}=\vec{p}_F/\vert \vec{p}_F \vert$, because excitations tangential to the Fermi surface are forbidden due to the Pauli principle.
The momentum of a low-energy mode near the Fermi surface can then be written as $\vec{p} \equiv \vec{p}_F+\vec{l}$, with $\vec{p}_F= \mu \vec{v}$, and a residual momentum $\vec{l}$, with $|\vec{l}\,| \lesssim |\vec{p}_F|$~\cite{Hong:1998tn, Hong:1999ru}. 
We then Fourier-decompose the quark spinors $\psi(x)$ in terms of Fermi-surface modes $\psi(\vec{v},x)$,
\begin{subequations}\label{decomp1}
\begin{align}
\psi(x) &= \sum_{\vec{v}} e^{i\mu \vec{v}\cdot \vec{x}}  \psi(\vec{v}, x) = \sum_{\vec{v}} e^{i\mu \vec{v}\cdot \vec{x}} \left[ \psi_{+}(\vec{v}, x) + \psi_{-}(\vec{v}, x) \right]\;,\\
\psi_{\pm}(\vec{v}, x)& = \int\frac{d^4 l}{(2\pi)^4} e^{-i l\cdot x} \psi_{\pm}(\vec{v}, l )\;.
\end{align}
\end{subequations}
where the sum over $\vec{v}$ runs over all possible values of the Fermi velocity on the Fermi surface and $\vert l_0 \vert \lesssim \mu$. 
The Fermi-surface mode $\psi(\vec{v}, x)$ has been decomposed into a slow $(+)$ and a fast $(-)$ eigenmode of the Hamiltonian $\mathcal{H}$ corresponding to the Lagrangian~\eqref{qcdl} in the non-interacting limit:
\begin{align}
\mathcal{H} \psi_{\pm}(\vec{v}, x) = \left( \vec{p}\cdot \vec{\alpha}- \mu  \right) \psi_{\pm}(\vec{v}, x) = E_{\pm}  \psi_{\pm}(\vec{v}, x) =\left( \pm \vert \vec{p}\,\vert - \mu  \right) \psi_{\pm}(\vec{v}, x)\;,
\end{align}
with $\vec{\alpha} = \gamma_0 \vec{\gamma}$. 
Each mode satisfies the following relations:
\begin{align}
\vec{p} \cdot \vec{\alpha}\, \psi_{\pm}(\vec{v}, x) = \pm \left\vert \vec{p}\right\vert \psi_{\pm}(\vec{v}, x)=\pm \left\vert \mu\vec{v}+\vec{l}_{\parallel}+\vec{l}_{\bot} \right\vert \psi_{\pm}(\vec{v}, x)\;,
\end{align}
where $\vec{l}_{\parallel}=\vec{v} (\vec{v}\cdot \vec{l})$ and $\vec{l}_{\bot}=\vec{l}-\vec{v} (\vec{v}\cdot \vec{l})$. 
In the limit where the momentum approaches the Fermi surface ($|\vec{l}\, | \ll |\vec{p}_F|$, $|l_0 | \ll \mu$), the quark modes are almost on-shell and the following projection becomes exact:
\begin{align}
\psi_{\pm}(\vec{v}, x) & \equiv P_{\pm}\psi(\vec{v}, x) \equiv \frac{1\pm \vec{\alpha}\cdot\vec{v}}{2}\psi(\vec{v}, x)\;.
\label{eq:fastslowmodes}
\end{align}
Since the surface modes have momenta $ \vert \vec{p}\,\vert \simeq \mu$, the slow mode $\psi_{+}(\vec{v}, x)$ requires a small excitation energy $|E_+| \ll \mu$, while the fast mode $\psi_{-}(\vec{v}, x)$ requires a large excitation energy $\vert E_{-} \vert \simeq 2\mu$.
Using the algebraic relations (2.9) of Ref.~\cite{Hong:1999ru} for the decomposition of the high-density QCD Lagrangian, i.e.,
\begin{align}
\bar{\psi}_{+}(\vec{v}, x)  P_{-} \gamma^{\mu} P_{+} \psi_{+}(\vec{v}, x)  &= V^{\mu}\bar{\psi}_{+}(\vec{v}, x) \gamma^{0}\psi_{+}(\vec{v}, x)\; ,\nonumber\\
\bar{\psi}_{-}(\vec{v}, x) P_{+} \gamma^{\mu} P_{-}\psi_{-}(\vec{v}, x) &=\bar{V}^{\mu}\bar{\psi}_{-}(\vec{v}, x) \gamma^{0}\psi_{-}(\vec{v}, x)\;,\nonumber\\
\bar{\psi}_{-}(\vec{v}, x)P_{+} \gamma^{\mu} P_{+} \psi_{+}(\vec{v}, x)& =\bar{\psi}_{-}(\vec{v}, x)\gamma_\bot^\mu\psi_{+}(\vec{v}, x)\;,\nonumber\\
\bar{\psi}_{+}(\vec{v}, x)P_{-} \gamma^{\mu} P_{-} \psi_{-}(\vec{v}, x)&=\bar{\psi}_{+}(\vec{v}, x)\gamma_\bot^\mu\psi_{-}(\vec{v}, x)\;,\label{algrel1}
\end{align}
where $V^\mu=(1,\vec{v})$, $\bar{V}^\mu=(1,-\vec{v})$, and $\gamma_\bot^\mu = \gamma^\mu -\gamma_\parallel^\mu$ with $\gamma_\parallel^\mu=(\gamma^0,\vec{v}
(\vec{v}\cdot \vec{\gamma}))$, the Lagrangian~\eqref{qcdl} can be expressed in terms of the slow and fast eigenmodes as
\begin{align}
\mathcal{L}_{\textrm{QCD}} & =  -\frac{1}{4}\mathcal{F}^{a}_{~\mu\nu}\mathcal{F}^{a\mu\nu}\nonumber\\
&\quad+\sum_{\vec{v}}\big[  \bar{\psi}_{+}(\vec{v}, x) \gamma^0(i V^{\mu} D_{\mu}) \psi_{+}(\vec{v}, x) +\bar{\psi}_{-}(\vec{v}, x) \gamma^0(2\mu + i \bar{V}^{\mu} D_{\mu}) \psi_{-}(\vec{v}, x) \nonumber\\
&\qquad\quad\quad+ \bar{\psi}_{+}(\vec{v}, x)\; i \,\slash \hspace{-0.25cm}
D_\bot \psi_{-}(\vec{v}, x) +\bar{\psi}_{-}(\vec{v}, x) \; i\, \slash \hspace{-0.25cm}
D_\bot \psi_{+}(\vec{v}, x)  \big]\;.~\label{dlagdkh}
\end{align}
The fast mode $\psi_{-}(\vec{v}, x)$ can be integrated out, which effectively corresponds to replacing it by using its equation of motion ($\delta \mathcal{S}_{\textrm{QCD}} / \delta \bar{\psi}_{-}(\vec{v}, x)  =0 $): 
\begin{align}
\psi_{-} (\vec{v}, x)=-\frac{\gamma^0}{2\mu+i\bar{V}^{\mu} D_{\mu}}\; i\,\slash
\hspace{-0.25cm} D_\bot \psi_{+}(\vec{v}, x)\;.~\label{fastmode}
\end{align}
Then, the high-density QCD Lagrangian~\eqref{qcdl} can be reduced to that of the so-called high-density effective theory (HDET)~\cite{Hong:1998tn, Hong:1999ru, Schafer:2003jn}:
\begin{align}
\mathcal{L}_{\textrm{QCD}} & \simeq -\frac{1}{4}\mathcal{F}^{a}_{~\mu\nu}\mathcal{F}^{a\mu\nu}+  \sum_{\vec{v}}\left[ \bar{\psi}_{+}(\vec{v}, x) \gamma^0(i V^{\mu} D_{\mu})
\psi_{+}(\vec{v}, x) - \bar{\psi}_{+}(\vec{v}, x)  \frac{\gamma^0}{2\mu+i\bar{V}^{\mu} D_{\mu}}
(\slash \hspace{-0.25cm}  D_\bot)^2 \psi_{+}(\vec{v}, x) \right]+\ldots\;.
\end{align}
One can calculate perturbatively the scattering processes of the relevant modes around a given small patch on the Fermi surface by using HDET.  
In the limit $\mu\gg \Lambda_{\textrm{QCD}}$, asymptotic freedom applies and the effective form of the four-quark interaction channels~\eqref{4qeff} can be constructed by summing up ladder diagrams and matching them to the corresponding channel of QCD. 
However, as non-perturbative effects such as instanton contributions may become important at  $\mu\simeq \mathcal{O}(\Lambda_{\textrm{QCD}})$, fermion contact interactions as considered in NJL-like models will be investigated as a first step.

\subsection{Surface zero-modes and instanton background}

\begin{figure}
\includegraphics[height=7cm]{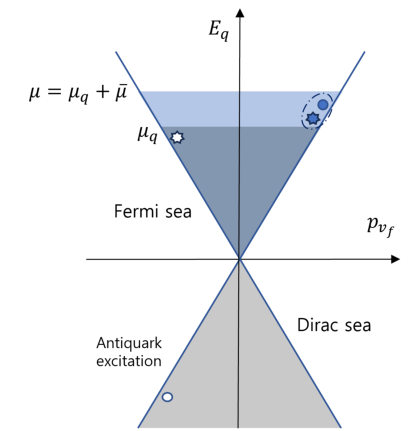} 
\caption{(Color online) The Dirac cone for a chemical potential $\mu  \neq 0$. 
A quark (filled circle) can be paired with hole state in the Dirac sea (empty circle) or the Fermi sea (empty star). 
Each hole state plays the role of an antiparticle state (filled star). 
An analogous, detailed discussion can be found in Ref.~\cite{Kojo:2009ha}. 
Note that the pairing occurs in a shell of thickness $\bar{\mu}$ around the Fermi surface, i.e., for chemical potentials $\in [\mu_q, \mu]$, where $\mu = \mu_q + \bar{\mu}$.}\label{fig1}
\end{figure}

\subsubsection{GN-like model and scalar condensation}

At intermediate densities, where $\mu\simeq \mathcal{O}(\Lambda_{\textrm{QCD}})$, it is not adequate to describe the dynamics in terms of perturbative HDET. 
However, the effective form of the marginal four-quark interaction can be expressed in terms of the relevant slow modes of HDET. 
In the following, we consider the scalar-isoscalar quark bilinear and its corresponding condensation pattern. 
With the help of Eq.~(\ref{decomp1}), the scalar-isoscalar quark condensate can be expressed as follows:
\begin{align}
\langle \bar{\psi} (x) \psi (x) \rangle = \sum_{\vec{v},\vec{v}'} e^{i\mu (\vec{v}-\vec{v}^{\, \prime})\cdot \vec{x}} \langle \bar{\psi}(\vec{v}^{\, \prime}, x) \psi (\vec{v}, x) \rangle\;.\label{qc}
\end{align}
The interaction between quarks becomes marginal only if the collinear scattering occurs between the slow and fast modes (e.g.,  quark and anti-quark states) at the same side of the Fermi surface, $\vec{v}\simeq  \vec{v}^{\, \prime}$, or between the slow modes (e.g., 
quark and hole states) at opposite sides of the Fermi surface,  $\vec{v}\simeq - \vec{v}^{\, \prime}$~\cite{Deryagin:1992rw}. 
If $\vec{v}\simeq  \vec{v}^{\, \prime}$, the respective contribution to Eq.~\eqref{qc} can be written as)
\begin{align}
&  \sum_{\vec{v}} \left[\langle \bar{\psi}_{+} (\vec{v}, x) \psi_{-} (\vec{v}, x) \rangle+\langle \bar{\psi}_{-} (\vec{v}, x) \psi_{+} (\vec{v}, x) \rangle \right]\;, \label{qc2}
\end{align}
because $\langle \bar{\psi}_{\pm} (\vec{v}, x) \psi_{\pm} (\vec{v}, x) \rangle=\langle \bar{\psi}_{\pm} (\vec{v}, x) P_{\mp} P_{\pm}\psi_{\pm} (\vec{v}, x) \rangle=0$. 
The remaining condensates in Eq.~\eqref{qc2} are quark-antiquark condensates,  where it costs at least an energy $\epsilon \simeq 2\mu$ to excite the fast mode. Therefore, the contribution from Eq.~\eqref{qc2} will be neglected in the following.
On the other hand, for the case of $\vec{v}\simeq  - \vec{v}^{\, \prime}$, the quark condensate becomes inhomogeneous and the respective contribution to Eq.\ (\ref{qc}) reads
\begin{align}
&  \sum_{\vec{v}} e^{2i\mu \vec{v} \cdot \vec{x}}  \left[\langle \bar{\psi}_{+} (-\vec{v}, x) \psi_{+} (\vec{v}, x) \rangle+\langle \bar{\psi}_{-} (-\vec{v}, x) \psi_{-} (\vec{v}, x) \rangle \right]  \;, \label{qc3}
\end{align}
since according to Eq.~(\ref{eq:fastslowmodes}), $\psi_\pm(-\vec{v},x) \equiv \psi_\mp (\vec{v}, x)$.
The remaining condensates cost minimal energy ($\epsilon \simeq 0$) for the excitation of the slow modes, and therefore $\langle \bar{\psi}(x) \psi(x) \rangle$ will be approximated by the term in Eq.\ (\ref{qc3}) in the following.

According to Eq.~(\ref{fastmode}), the fast modes are suppressed by a factor $\sim \mathcal{O}(1/\mu)$ compared to the slow modes. 
Therefore, the fast-mode condensate in Eq.~(\ref{qc3}) is suppressed by a factor $\sim \mathcal{O}(1/\mu^2)$ compared to the slow-mode condensate, and can be neglected. 
The inhomogeneous quark condensate can be expressed in terms of quark and hole modes with opposite Fermi velocities: 
\begin{align}
\langle \bar{\psi} (x) \psi (x) \rangle =  \sum_{\vec{v}} e^{2i\mu \vec{v} \cdot \vec{x}} \langle \bar{\psi}_{+} (-\vec{v}, x) \psi_{+} (\vec{v}, x) \rangle \quad \Longrightarrow \quad  4 \pi \cos(2\mu \vert \vec{x} \vert) \sigma \mathit{\Delta}_{\bot}\;, \label{qc4}
\end{align}
where we assumed that the system is isotropic, such that $\langle \bar{\psi}_{+} (-\vec{v}, x) \psi_{+} (\vec{v}, x) \rangle=\langle \bar{\psi}_{+} (\vec{v}, x) \psi_{+} (-\vec{v}, x) \rangle=\sigma\mathit{\Delta}_{\bot}$. 
Here, $\mathit{\Delta}_{\bot}$ denotes the mean contribution from a small transverse patch on the Fermi surface (with dimension $\textrm{MeV}^2$), while $\sigma$ is the condensation amplitude of the longitudinal modes (with dimension MeV). 
Adding a condensate in the pseudoscalar channel, this form can be cast into a chiral density-wave form.

In the Wilsonian scaling of the fermion action to the Fermi surface, the collinear four-quark interactions with minimal momentum transfer ($\vec{p}_1 \simeq \pm \vec{p}_2$, $\vert \vec{p}_i \vert =\vert \vec{p}_F \vert$) becomes marginal (see Section 1.3 of Ref.~\cite{Nardulli:2002ma} for details). 
Thus, if one considers the effective form of the scalar interaction given in Eq.~\eqref{4qeff}, for the Overhauser-type pairing the Fermi velocities assigned to the quark fields should fulfill $\vec{v}_1 \simeq -\vec{v}_3 \simeq -\vec{v}_2 \simeq \vec{v}_4$. 
Therefore, one can make the following Ansatz for the scalar four-quark interaction channel in a GN-like model, where the quarks couple to the scalar-meson state:
\begin{align}
\mathcal{L}_{4qs}&= \left[ \bar{\psi} (x) \psi (x) \right]^2 \Longrightarrow  \sum_{\vec{v}} \frac{1}{\mu^2} \frac{g_s}{n_D} \left[  \bar{\psi}_{+} (-\vec{v}, x) \psi_{+} (\vec{v}, x)  \right] \left[  \bar{\psi}_{+} (\vec{v}, x) \psi_{+} (-\vec{v}, x)  \right]\;, \label{4qs}
\end{align}
where $n_D \equiv N_fN_c$ denotes the degeneracy factor and the phase factors $ e^{\pm 2i\mu \vec{v} \cdot \vec{x}}$ cancel between the two terms in Eq.\ (\ref{4qs}). 
The four-fermion interaction strength in the scalar channel is denoted as $g_s$.
One can find the explicit quark-scalar meson coupling channel through bosonizing the four-quark interaction via a Hubbard-Stratonovich transformation~\cite{1957SPhD....2..416S, Hubbard:1959ub}.
To  make the connection of this channel with QCD, one should consider the surface zero-mode correlation in an instanton-background field, which is subject of the next subsection.

\subsubsection{HDET and the surface zero-modes}

Instantons are topologically non-trivial solutions of the classical Yang-Mills equations of motion in Euclidean space~\cite{tHooft:1976rip, Belavin:1975fg}. 
These special configurations minimize the gauge action and correlate topologically different vacua, where the number of left- and right-handed quark zero modes ($I=n_R-n_L$) differs by an integer (details are given in Appendix \ref{app:B}).

The QCD Lagrangian~\eqref{qcdl} in the presence of a quark-number chemical potential $\mu$ can be written in Euclidean space as
\begin{align}
\mathcal{L}^{\textrm{E}}_{\textrm{QCD}} & =  \frac{1}{4}\mathcal{F}^{a}_{\mu\nu}\mathcal{F}^{a}_{\mu\nu}- \bar{\psi} (i \, \slash \hspace{-0.25cm}  D_{\textrm{E}} +
\mu \gamma^0) \psi\;,\label{qcdle}
\end{align}
where the covariant derivative in Euclidean space is $D_{\textrm{E} \mu}=\partial_{\mu}-ig A_{\mu}$. 
The instanton field-strength tensor will be denoted as $\mathcal{G}_{\mu\nu}$.  
From the matter part of the Lagrangian~\eqref{qcdle}, one can define a $\mu$-dependent Dirac operator as $i \, \slash \hspace{-0.25cm}  D_{\textrm{E}} (\mu) \equiv i \slash \hspace{-0.25cm}  D_{\textrm{E}} +\mu \gamma^0$, which has following properties:
\begin{align}
i \, \slash \hspace{-0.25cm}  D_{\textrm{E}} (\mu)  \psi_\lambda  = \lambda(\mu)\psi_\lambda \;, \quad \left[i \,\slash \hspace{-0.25cm}  D_{\textrm{E}} (\mu)\right]^{\dagger} = -i \, \slash \hspace{-0.25cm}  D_{\textrm{E}} (-\mu)\;, \quad \lambda^*(\mu)=- \lambda(-\mu)\;, \label{diracoprels}
\end{align}
where the gluon field in the covariant derivative corresponds to the instanton background field.
Since $\{ i \, \slash \hspace{-0.25cm}  D_{\textrm{E}} (\mu), \gamma_5 \}=0$,  the eigenmodes $\psi_{\lambda}$ always exist in pairs with $\psi_{-\lambda }$ ($\lambda \neq0$). 
In the vacuum limit ($\mu \rightarrow 0$), $i \, \slash \hspace{-0.25cm}  D_{\textrm{E}} ( 0)$ becomes anti-Hermitian. 
Due to the definite anti-hermiticity of $i \,\slash \hspace{-0.25cm}  D_{\textrm{E}} ( 0)$, the eigenmodes satisfy an orthonormality relation, and one can easily show from the eigenmode decomposition of the quark propagator in an instanton background that only zero modes contribute to the scalar quark condensate~\eqref{qc2} (see Appendix~\ref{appc} for details). 
 At finite density ($\mu\simeq O(\Lambda_{\textrm{QCD}})$), the definite \textcolor{violet}{anti-}hermiticity of the Dirac operator is lost and one should find a new set of basis states for the orthonormality relation~\cite{Bruckmann:2013rpa}:
\begin{align}
i \,\slash \hspace{-0.25cm}  D_{\textrm{E}} (\mu)  \psi_{m}(\mu)  = \lambda_{m}(\mu)\psi_{m}(\mu)\;, \quad
\psi^{\dagger}_{n}(-\mu) i\, \slash \hspace{-0.25cm}  D_{\textrm{E}} (\mu)    = \lambda_{n}(\mu) \psi^{\dagger}_{n}(-\mu)\;, \quad
\textrm{STr}\left[ \psi^{\dagger}_{n}(-\mu)\psi_{m}(\mu)\right]=\delta_{nm}\;,
\label{eq:bi-ortho}
\end{align}
where `STr' denotes the supertrace including a summation over color-spin indices  and a space-time integration. 
From the bi-orthonormal relationship (\ref{eq:bi-ortho}) between $\psi^{\dagger}_{n}(-\mu)$ and $\psi_{m}(\mu)$, one expects that the quark zero modes for the scalar condensation should be found in pairs for $\pm\mu$ and the proper index theorem can be written as follows~\cite{Bruckmann:2013rpa, Kanazawa:2011tt}:
\begin{align}
I=n_R(\mu)-n_L(-\mu)=n_R(-\mu)-n_L(\mu)\;.\label{index2}
\end{align}
As $\psi_{\lambda}(-\mu)$ can be understood as an antiparticle mode of the system with chemical potential $\mu$, one may recall the decomposition~\eqref{decomp1} and investigate the quark field again.

The standard decomposition of the quark field in terms of annihilation operators $b_{s} ( \vec{p} )$ and antiquark creation operators $d^{\dagger}_{s}( \vec{p}) $ for (anti-)quarks with spin $s$ and momentum $\vec{p}$ reads
\begin{align}
\psi(x) &= \sum_{s} \int \frac{d^4p}{(2\pi)^4}\left[ b_{s}(\vec{p}) u_{s}(\vec{p})e^{-ip\cdot x}+  d^{\dagger}_{s}(\vec{p}) v_{s}(\vec{p})e^{ip\cdot x}  \right]
\;.\label{qexp1}
\end{align}
This decomposition can be mapped to the decomposition (\ref{decomp1}), noting that, at finite chemical potential, the energy $p^0$ of a particle mode is measured relative to $\mu$, such that $p^0 \equiv l^0$.
Furthermore, employing $\vec{p} = \mu \vec{v} + \vec{l}$, we immediately deduce that $\int d^4p \equiv \sum_{\vec{v}} \int d^4l$. 
A direct comparison of Eqs.~(\ref{decomp1}) and (\ref{qexp1}) then yields
\begin{subequations}\label{fmode}
\begin{align}
\psi_{+}(\vec{v}, x) &= \sum_{s} \int \frac{d^4l}{(2\pi)^4}  b_{s} ( \vec{v}, l) u_{s}( \vec{v}, l)e^{-il\cdot x}\;, \\
\psi_{-}(\vec{v}, x) &=   \sum_{s} \int \frac{d^4l}{(2\pi)^4}   d^{\dagger}_{s}( \vec{v}, -l) v_{s}( \vec{v},- l) e^{-2i\mu \vec{v}\cdot \vec{x}}e^{-il\cdot  x}\;. \label{fmodeb}
\end{align}
\end{subequations}
The origin of the $2\mu$-scale of the fast modes is obvious from Eq.~\eqref{fmodeb}. 

The matter part of the decomposition~\eqref{dlagdkh} can be rewritten in the following matrix form:
\begin{align}
\mathcal{L}^{\textrm{E}}_{\bar{\psi}D \psi} & = -\sum_{\vec{v}} \left(  
\bar{\psi}_{-} (\vec{v}, x) \, , \;
  \bar{\psi}_{+} (\vec{v}, x) \right)\left( {\begin{array}{cc}
 i \,\slash \hspace{-0.25cm}  D_{\textrm{E}\bot} &i \,\slash \hspace{-0.25cm}  D_{\textrm{E}  \parallel}  + 2\mu\gamma^0\\
i \,\slash \hspace{-0.25cm}  D_{\textrm{E} \parallel}  &  i\, \slash \hspace{-0.25cm}  D_{\textrm{E} \bot}\\
  \end{array} } \right)  \left( {\begin{array}{c}
\psi_{+} (\vec{v}, x)\\
  \psi_{-} (\vec{v}, x)  \end{array}  }\right)\;, \label{eq:22}
\end{align}
where $i \,\slash \hspace{-0.25cm}  D_{\textrm{E}\parallel} \psi_{+} (\vec{v}, x) = \gamma^0 ( iV _{\mu} D_{\textrm{E}\mu})  \psi_{+} (\vec{v}, x) $ and $i \,\slash \hspace{-0.25cm}  D_{\textrm{E}\parallel} \psi_{-} (\vec{v}, x) = \gamma^0  (i\bar{V} _{\mu} D_{\textrm{E}\mu})  \psi_{-} (\vec{v}, x) $.
For the sake of convenience, we abbreviate the $8 \times 8$ matrix in Eq.\ (\ref{eq:22}) by $i \, \slash \hspace{-0.25cm}  D_{\textrm{E}} (\mu)$ in the following algebraic calculations. 
Since the zero modes satisfy $i \,\slash \hspace{-0.25cm}  D_{\textrm{E}} (\mu) \psi_0 (\vec{v}) = 0$, the relation $[i \,\slash \hspace{-0.25cm}  D_{\textrm{E}} (\mu)]^2  \psi_0  (\vec{v})= 0$ reads:
\begin{align}
  &\left( {\begin{array}{cc}
(i \,\slash \hspace{-0.25cm}  D_{\textrm{E} \parallel} +2\mu \gamma^0 ) i \,\slash \hspace{-0.25cm}  D_{\textrm{E} \parallel} +(i \,\slash \hspace{-0.25cm}  D_{\textrm{E} \bot})^2 &(i \,\slash \hspace{-0.25cm}  D_{\textrm{E} \parallel}+2\mu \gamma^0 ) i \,\slash \hspace{-0.25cm}  D_{\textrm{E} \bot}  + i \,\slash \hspace{-0.25cm}  D_{\textrm{E}\bot} (i \,\slash \hspace{-0.25cm}  D_{\textrm{E}\parallel} +2\mu \gamma^0 ) \\
 i \,\slash \hspace{-0.25cm}  D_{\textrm{E} \bot}\; i \,\slash \hspace{-0.25cm}  D_{\textrm{E} \parallel}  +  i\, \slash \hspace{-0.25cm}  D_{\textrm{E}\parallel}\; i \, \slash \hspace{-0.25cm}  D_{\textrm{E} \bot}&  i \,\slash \hspace{-0.25cm}  D_{\textrm{E}\parallel}   (i \,\slash \hspace{-0.25cm}  D_{\textrm{E}\parallel} +2\mu \gamma^0 ) +(i \,\slash \hspace{-0.25cm}  D_{\textrm{E} \bot})^2\\
  \end{array} } \right)  \left( {\begin{array}{c}
\psi_{+0} (\vec{v}, x)\\
  \psi_{-0} (\vec{v}, x)  \end{array}  }\right)\nonumber\\
 &\simeq    \left( {\begin{array}{cc}
(i \,\slash \hspace{-0.25cm}  D_{\textrm{E}\parallel} +2\mu \gamma^0 ) i \,\slash \hspace{-0.25cm}  D_{\textrm{E}\parallel} +(i \,\slash \hspace{-0.25cm}  D_{\textrm{E}\bot})^2 &0\\
0&  i \,\slash \hspace{-0.25cm}  D_{\textrm{E}\parallel} (i \,\slash \hspace{-0.25cm}  D_{\textrm{E} \parallel} +2\mu \gamma^0 ) +(i \,\slash \hspace{-0.25cm}  D_{\textrm{E}\bot})^2\\
  \end{array} } \right)  \left( {\begin{array}{c}
\psi_{+0} (\vec{v}, x)\\
  \psi_{-0} (\vec{v}, x)  \end{array}  }\right)=0\;, \label{decompdops1}
\end{align}
where we have used $\psi_0 (\vec{v})= \left( {\begin{array}{c}
\psi_{+0} (\vec{v}, x)\\
  \psi_{-0} (\vec{v}, x)  \end{array}  }\right)$. 
  The second line of Eq.\ (\ref{decompdops1}) follows from the fact that the off-diagonal elements are suppressed by $  \mathcal{O}(1/\mu)$, since 
\begin{subequations}\label{offdiag}
\begin{align}
 (i \,\slash \hspace{-0.25cm}  D_{\textrm{E} \bot}\; i \,\slash \hspace{-0.25cm}  D_{\textrm{E} \parallel}  +  i\, \slash \hspace{-0.25cm}  D_{\textrm{E}\parallel}\; i \, \slash \hspace{-0.25cm}  D_{\textrm{E} \bot})\psi_{+ 0} (\vec{v}, x) = ig \gamma_{\bot \beta} \gamma_{\parallel \alpha} \mathcal{G}_{\alpha\beta}\psi_{+ 0} (\vec{v}, x)=igV_{\bot \beta} V_{\alpha}\mathcal{G}_{\alpha\beta}\psi_{+ 0} (\vec{v}, x)\;,\nonumber\\
 (i \,\slash \hspace{-0.25cm}  D_{\textrm{E} \bot}\; i \,\slash \hspace{-0.25cm}  D_{\textrm{E} \parallel}  +  i\, \slash \hspace{-0.25cm}  D_{\textrm{E}\parallel}\; i \, \slash \hspace{-0.25cm}  D_{\textrm{E} \bot})\psi_{- 0} (\vec{v}, x) = ig \gamma_{\bot \beta} \gamma_{\parallel \alpha} \mathcal{G}_{\alpha\beta}\psi_{- 0} (\vec{v}, x)=ig \bar{V}_{\bot \beta} \bar{V}_{\alpha}\mathcal{G}_{\alpha\beta}\psi_{- 0} (\vec{v}, x)\;, 
\end{align}
\end{subequations}
where $V_{\bot \beta}=(0, \vec{l}_{\bot}/\vert \mu\vec{v}+\vec{l} \vert)$ and $\bar{V}_{\bot \beta}=(0, -\vec{l}_{\bot}/\vert \mu\vec{v}+\vec{l} \vert)$. 
Each element of the column vector resulting from the last line of Eq.~\eqref{decompdops1} can be simply expressed as
\begin{align}
 \left[ 2\mu (iV _{\mu} D_{\textrm{E}\mu}) + (i \,\slash \hspace{-0.25cm}  D_{\textrm{E}\parallel})^2+ (i \,\slash \hspace{-0.25cm}  D_{\textrm{E} \bot})^2 \right] \psi_{\pm0} (\vec{v}, x) \simeq  \left[ 2\mu (iV _{\mu} D_{\textrm{E}\mu}) + D_{\textrm{E}}^2+\frac{g}{2} \frac{1\pm'\gamma_5}{2}\sigma_{\mu\nu}\mathcal{G}_{\mu\nu} \right]\psi_{\pm0} (\vec{v}, x)=0\;,\label{instzero}
\end{align}
where $\sigma_{\alpha \beta}\gamma_5 = \frac{1}{2} \epsilon_{\alpha\beta\mu\nu}\sigma_{\mu\nu}$ and the following (anti-)self-duality relation of the (anti-)instanton field-strength tensor has been used:
\begin{align}
\mathcal{G}_{\mu\nu}=\pm' \frac{1}{2} \epsilon_{\mu\nu\rho\sigma}\mathcal{G}_{\rho\sigma} \;.\label{insts}
\end{align}
Here, $\pm'$ denotes the instanton and anti-instanton field strength, respectively (while the $\pm$ at $\psi_{\pm 0}(\vec{v},x)$ denotes the slow/fast mode). 
Since we are investigating phenomena near the Fermi surface,
$\vert D_{\textrm{E}\mu} \vert \lesssim \Lambda_{\textrm{QCD}}$ by assuming a weak coupling to the instanton background.
In the vacuum limit $\mu \rightarrow 0$,  $2\mu (iV_{\mu} D_{\textrm{E}\mu}) + D_{\textrm{E}}^2 \rightarrow D_{\textrm{E}}^2$  becomes negative definite. 
Since for left-handed modes the right-handed projector $(1 + \gamma_5)/2$ eliminates the last term in Eq.~\eqref{instzero}, while the left-handed projector $(1 - \gamma_5)/2$  does the same for right-handed modes, the amplitude of the left-handed (right-handed) zero-mode $\psi_{L(R)0}(\vec{v}, x)$ is required to vanish. 
However, the right-handed (left-handed) zero mode $\psi_{R(L)0}(\vec{v}, x)$ does not need to vanish, meaning that there is scalar quark condensation which correlates topologically different vacua ($I=\pm 1$). 
As long as we are in the low-$\mu$ region, this conclusion will remain valid, i.e., scalar quark condensation of the quark and antiquark modes via instanton interactions persist.

\begin{figure}
\includegraphics[height=5.5cm]{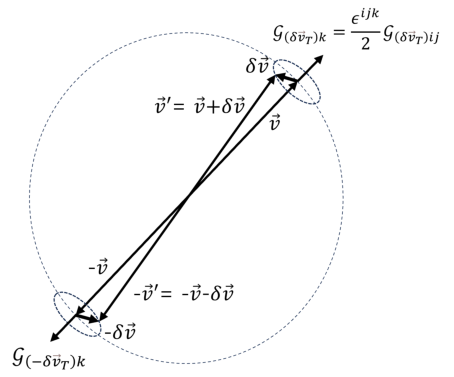} 
\caption{A simple diagram depicting the transverse perturbation in a small patch on the Fermi surface. 
The correlation of the instanton field strength on both sides of the Fermi sphere can correlate the quark and hole state around the patch in both directions of a given Fermi velocity.}\label{fig2}
\end{figure}

On the other hand, if one considers a state $\psi_{\lambda}(\mu)$ near the Fermi surface with a given Fermi velocity $\vec{v}$, another candidate for $\psi_{\lambda}(-\mu)$ is a quark state on the other side of the Fermi surface, in the direction $-\vec{v}$, since it also costs an energy $2\mu$ to flip a surface quark state to the opposite side of Fermi sphere. 
The antiparticle creation operator $d^{\dagger}_{s}( \vec{v}, -l)$ in the fast mode~\eqref{fmode} is indeed the annihilation operator acting on an occupied state ($-s,  -\mu\vec{v}+\vec{l}$) in the Dirac sea. 
If one separates the contribution $\sim d^{\dagger}_{s}( \vec{v}, -l)$ by a proper effective approach, then the separated slow mode describes the $d$-type particle state obtained by annihilation of the ($-s,  -\mu\vec{v}+\vec{l}$) state below the reference ground level where the effective dynamics is defined. 
From a technical point of view, this operator can describe the hole state obtained by annihilation of the ($-s,  -\mu\vec{v}+\vec{l}$) state in the Fermi sphere if the reference ground level is set as the Fermi surface in the direction $-\vec{v}$.
By denoting $d^{\dagger}_{s}( \vec{v}, -l)$ as $b_{-s}(-\vec{v}, l)$, which annihilates an occupied state  ($-s,  -\mu\vec{v}+\vec{l}$) in the Fermi sea, the $d$-type slow modes can be utilized to describe dynamics of the $b$-type slow modes at the opposite side of the Fermi sphere. 
One can decompose the quark field in a symmetric way to obtain the second version of the effective description for the relevant slow modes moving in both directions of a given Fermi velocity. If one writes the quark field as 
\begin{align}
\psi(x) &= \sum_{s} \int \frac{d^4p}{(2\pi)^4}\left[ b_{s}(\vec{p}) u_{s}(\vec{p})e^{-ip\cdot x}+  d^{\dagger}_{s}(\vec{p}) v_{s}(\vec{p})e^{ip\cdot x}  \right]\quad \Longrightarrow\quad \sum_{\vec{v}}\left[  e^{i\mu \vec{v}\cdot \vec{x}}\psi_{p}(\vec{v}, x) +   e^{-i\mu \vec{v}\cdot \vec{x}}\psi_{n}(\vec{v}, x)  \right]\;,
\end{align}
the positive- and negative-energy modes with residual momentum $\vert l_{\mu} \vert \lesssim \mu$ can be identified as follows:
\begin{align}
\psi_{p}(\vec{v}, x) &= \sum_{s} \int \frac{d^4l}{(2\pi)^4}  b_{s} ( \vec{v}, l ) u_{s}( \vec{v}, l ) e^{-il\cdot x}\;,\\
\psi_{n}(\vec{v}, x) &=   \sum_{s} \int \frac{d^4l}{(2\pi)^4}   d^{\dagger}_{s}( \vec{v}, -l) v_{s}( \vec{v}, -l)e^{-il\cdot x}\;.
\end{align}
 Then, one can express the matter part of the Lagrangian density~\eqref{qcdle} in Euclidean space as follows:
\begin{align}
\mathcal{L}^{\textrm{E}}_{\bar{\psi}D \psi} & =-\sum_{\vec{v}}\big[  \bar{\psi}_{p}(\vec{v}, x) \gamma^0(i V_{\mu} D_{\textrm{E}\mu}) \psi_{p}(\vec{v}, x) +\bar{\psi}_{n}(\vec{v}, x) \gamma^0( i \bar{V}_{\mu} D_{\textrm{E}\mu}) \psi_{n}(\vec{v}, x) \nonumber\\
&\qquad\quad+e^{-2i\mu \vec{v}\cdot \vec{x}} \bar{\psi}_{p}(\vec{v}, x)\;i \, \slash \hspace{-0.25cm}
D_{\textrm{E}\bot} \psi_{n}(\vec{v}, x) +e^{2i\mu \vec{v}\cdot \vec{x}}\bar{\psi}_{n}(\vec{v}, x) \;i \,\slash \hspace{-0.25cm}
D_{\textrm{E}\bot} \psi_{p}(\vec{v}, x)  \big]\;,\label{dlagitp}
\end{align}
where the algebraic relations~\eqref{algrel1} have been used. This matter part can be reduced further to the effective form:
\begin{align}
\mathcal{L}^{\textrm{E}}_{\bar{\psi}D \psi} & =-\sum_{\vec{v}}\Bigg[ \bar{\psi}_{p}(\vec{v}, x) \gamma^0(i V_{\mu} D_{\textrm{E}\mu})
\psi_{p}(\vec{v}, x) - \bar{\psi}_{p}(\vec{v}, x)  \frac{\gamma^0}{2\mu+i\bar{V}_{\mu} D_{\textrm{E}\mu}}
(\slash \hspace{-0.25cm}  D_{\textrm{E}\bot})^2 \psi_{p}(\vec{v}, x) \nonumber\\
&\qquad\qquad \quad+\bar{\psi}_{n}(\vec{v}, x) \gamma^0(i \bar{V}_{\mu} D_{\textrm{E}\mu})
\psi_{n}(\vec{v}, x) - \bar{\psi}_{n}(\vec{v}, x)  \frac{\gamma^0}{2\mu+iV_{\mu} D_{\textrm{E}\mu}}
(\slash \hspace{-0.25cm}  D_{\textrm{E}\bot})^2 \psi_{n}(\vec{v}, x) \Bigg]\;, \label{hdet2}
\end{align}
where the following equations of motion $\delta \mathcal{S}_{\textrm{QCD}}/ \delta \bar{\psi}_{p/n}(\vec{v}, x)=0 $ were used:
\begin{align}
i\, \slash \hspace{-0.25cm}  D_{\textrm{E} \parallel} \psi_{n} (\vec{v}, x)&=- e^{2i\mu \vec{v}\cdot \vec{x}} i\,\slash
\hspace{-0.25cm} D_{\textrm{E}\bot} \psi_{p}(\vec{v}, x)\;,\nonumber\\
i \,\slash \hspace{-0.25cm}  D_{\textrm{E} \parallel} \psi_{p} (\vec{v}, x)&=-e^{-2i\mu \vec{v}\cdot \vec{x}} i\,\slash
\hspace{-0.25cm} D_{\textrm{E}\bot} \psi_{n}(\vec{v}, x)\;.\label{eqm2}
\end{align}
From the effective Lagrangian~\eqref{hdet2}, one can see that the negative-energy slow mode $\psi_{n}(\vec{v}, x)$ follows the same longitudinal dynamics as the positive-energy slow mode moving in the opposite direction $\psi_{p}(-\vec{v}, x)$, since $\bar{V}_{\mu}=(i,-\vec{v})$. 
Considering that the $d$-type particle state with quantum numbers ($s,  \vec{v}, -l$) is obtained by annihilating the occupied ($-s,  -\vec{v}, l$) state below the ground level, one may match $\psi_{n}(\vec{v}, x)$ as follows:
\begin{align}
\psi_{n}(\vec{v}, x) &=   \sum_{s} \int \frac{d^4l}{(2\pi)^4}   d^{\dagger}_{s}( \vec{v}, -l) v_{s}( \vec{v}, -l)e^{-il\cdot x} \quad \Longrightarrow \quad \psi_{p}(-\vec{v}, x)=\sum_{s} \int \frac{d^4l}{(2\pi)^4}   b_{-s}( -\vec{v}, l) u_{-s}( -\vec{v}, l)e^{-il\cdot x}\;.
\end{align}
Thus, the conjugated field $\bar{\psi}_{n}(\vec{v}, x)$ describes the same dynamics as a hole state in the Fermi sea in the $-\vec{v}$ direction ($\bar{\psi}_{p}(-\vec{v}, x)$) and the condensate $\left\langle \bar{\psi}_{+}(-\vec{v}, x)\psi_{+}(\vec{v}, x) \right\rangle$ can be investigated in the $\{ \psi_{p}(\vec{v}, x), \psi_{n}(\vec{v}, x)\}$ basis. 
By denoting $\psi_{n}(\vec{v}, x)$  as $\psi_{+}(-\vec{v}, x)$, the matrix form of the matter part~\eqref{dlagitp} can be expressed accordingly as
\begin{align}
\mathcal{L}^{\textrm{E}}_{\bar{\psi}D \psi} 
 & = -\sum_{\vec{v}} \left(  
\bar{\psi}_{+} (-\vec{v}, x) \,, \;
  \bar{\psi}_{+} (\vec{v}, x) \right)\left( {\begin{array}{cc}
 e^{2i\mu \vec{v}\cdot \vec{x}} i \,\slash \hspace{-0.25cm}  D_{\textrm{E} \bot} &i \,\slash \hspace{-0.25cm}  D_{\textrm{E}\parallel}   \\
i\, \slash \hspace{-0.25cm}  D_{\textrm{E}\parallel}  & e^{-2i\mu \vec{v}\cdot \vec{x}} i \, \slash \hspace{-0.25cm}  D_{\textrm{E} \bot}\\
  \end{array} } \right)  \left( {\begin{array}{c}
\psi_{+} (\vec{v}, x)\\
  \psi_{+} (-\vec{v}, x)  \end{array}  }\right)\;,
\end{align}
and $[i \, \slash \hspace{-0.25cm}  D_{\textrm{E}} (\mu)]^2  \psi_0 = 0$ can be reduced as follows:
\begin{align}
\left( {\begin{array}{cc}
(i\, \slash \hspace{-0.25cm} D_{\textrm{E}\parallel} )^2 +e^{2i\mu \vec{v}\cdot \vec{x}} i \,\slash \hspace{-0.25cm} D_{\textrm{E}\bot} e^{2i\mu \vec{v}\cdot \vec{x}} i \, \slash \hspace{-0.25cm} D_{\textrm{E}\bot}   & e^{2i\mu \vec{v}\cdot \vec{x}}i \,\slash \hspace{-0.25cm}  D_{\textrm{E}\bot} \; i \,\slash \hspace{-0.25cm} D_{\textrm{E}\parallel}  + i\, \slash \hspace{-0.25cm} D_{\textrm{E}\parallel} e^{-2i\mu \vec{v}\cdot \vec{x}}i \,\slash \hspace{-0.25cm}  D_{\textrm{E}\bot}  \\[0.1cm]
e^{-2i\mu \vec{v}\cdot \vec{x}}i \,\slash \hspace{-0.25cm}  D_{\textrm{E}\bot} \; i \,\slash \hspace{-0.25cm} D_{\textrm{E}\parallel}  + i\, \slash \hspace{-0.25cm} D_{\textrm{E}\parallel} e^{2i\mu \vec{v}\cdot \vec{x}}i\, \slash \hspace{-0.25cm}  D_{\textrm{E}\bot}  & (i \,\slash \hspace{-0.25cm} D_{\textrm{E}\parallel} )^2 +e^{-2i\mu \vec{v}\cdot \vec{x}}i \,\slash \hspace{-0.25cm} D_{\textrm{E}\bot} e^{-2i\mu \vec{v}\cdot \vec{x}}i\, \slash \hspace{-0.25cm} D_{\textrm{E}\bot}  \\
  \end{array} } \right)  \left( {\begin{array}{c}
\psi_{+0} (\vec{v}, x)\\
  \psi_{+0} (-\vec{v}, x)  \end{array}  }\right)\nonumber\\
 \simeq   \left( {\begin{array}{cc}
(i \,\slash \hspace{-0.25cm}  D_{\textrm{E}\parallel} )^2  - i\, \slash \hspace{-0.25cm}  D_{\textrm{E} \parallel} \;i \, \slash \hspace{-0.25cm}  D_{\textrm{E} \bot} \frac{\gamma^0}{i \,\slash \hspace{-0.20cm}  D_{\textrm{E} \parallel} +2\mu \gamma^0} i \,\slash \hspace{-0.25cm}  D_{\textrm{E} \bot} &0\\
0&  (i \,\slash \hspace{-0.25cm}  D_{\textrm{E} \parallel} )^2  - i \, \slash \hspace{-0.25cm}  D_{\textrm{E} \parallel} \;i \, \slash \hspace{-0.25cm}  D_{\textrm{E} \bot} \frac{\gamma^0}{i \,\slash \hspace{-0.20cm}  D_{\textrm{E}\parallel} +2\mu \gamma^0} i \,\slash \hspace{-0.25cm}  D_{\textrm{E} \bot}\\
  \end{array} } \right)  \left( {\begin{array}{c}
\psi_{+0} (\vec{v}, x)\\
  \psi_{+0} (-\vec{v}, x)  \end{array}  }\right)=0\;,  \label{decompdops2}
\end{align}
where the off-diagonal elements are  suppressed as $\mathcal{O}(1/\mu)$ by the relation~\eqref{offdiag} and the constraints~\eqref{eqm2}.
The last line of Eq.~\eqref{decompdops2} can be simplified as
\begin{align}
  &\left[ (i\, \slash \hspace{-0.25cm} D_{\textrm{E}\parallel} )^2  - i \, \slash \hspace{-0.25cm}  D_{\textrm{E}\parallel} \;i \,\slash \hspace{-0.25cm}  D_{\textrm{E} \bot} \frac{\gamma^0}{i \,\slash \hspace{-0.25cm}  D_{\textrm{E} \parallel} +2\mu \gamma^0} i \,\slash \hspace{-0.25cm}  D_{\textrm{E} \bot} \right] \psi_{+0} (\pm \vec{v}, x) \nonumber\\
 &=  \left[ D_{\textrm{E} \parallel}^2+\frac{g}{2} \sigma_{\mu\nu}\mathcal{G}_{\parallel\mu\nu} + \frac{ i \, \slash \hspace{-0.25cm}  D_{\textrm{E}\parallel}}{i \,\slash \hspace{-0.25cm}  D_{\textrm{E}\parallel} +2\mu \gamma^0}\left(  D_{ \bot}^2+\frac{g}{2} \sigma_{\mu\nu}\mathcal{G}_{\bot\mu\nu}\right)\right] \psi_{+0} (\pm \vec{v}, x) =0\;.\label{dopeq2}
\end{align}
As seen in the fast-slow mode condensation case, an isotropic scalar quark condensate can be formed only in the vacuum limit $\mu \rightarrow 0$ (the negative slow mode becomes the negative-energy mode in vacuum) since the self-dual instanton field strength~\eqref{insts} guarantees the vanishing amplitude of $\psi_{L0}$. 
In the limit that the momentum goes to the Fermi surface with $\mu \gg \Lambda_{\textrm{QCD}}$, Eq.~\eqref{dopeq2} reduces to $\left( D_{\textrm{E}\parallel}^2+\frac{g}{2} \sigma_{\mu\nu}\mathcal{G}_{\parallel\mu\nu} \right) \psi_{+0} (\pm \vec{v}, x) =0$. 
Although the isotropic quark condensation is not possible at $\mu\simeq \mathcal{O}(\Lambda_{\textrm{QCD}})$, the velocity-oriented quark and hole condensation along a given Fermi-velocity axis is possible. 
Consider a small patch on the Fermi surface whose area is $\sim \mathcal{O}(l_{\bot}^2) < \Lambda_{\textrm{QCD}}^2 $ and a small transverse perturbation in the patch: $\vec{v}'=\vec{v}+\delta\vec{v} $ with $ \vert \delta \vec{v} \vert \equiv l_{\bot}/\mu \ll 1 $ and $\vec{v}\cdot \delta \vec{v} = 0$ (Fig.~\ref{fig2}). Then, $i \,\slash \hspace{-0.25cm}  D_{\textrm{E} \parallel}$  can be expressed as 
\begin{align}
  i\,  \slash \hspace{-0.25cm}  D_{\textrm{E}\parallel}  &\simeq i \left[\gamma_{4}D_{\textrm{E}4} +  (\vec{v}\cdot\vec{\gamma})\vec{v}\cdot \vec{D}_{\textrm{E}} +  (\delta \vec{v}_T \cdot \vec{\gamma}) \delta \vec{v}_T \cdot \vec{D}_{\textrm{E}} \right]\;,\\
   (i \,\slash \hspace{-0.25cm}  D_{\textrm{E}\parallel})^2  &\simeq  D_{\vec{v}}^2 + D_{\vec{v}_T}^2+\frac{g}{2} \sigma_{\mu\nu}\left (\mathcal{G}_{\vec{v} \mu\nu}+\mathcal{G}_{\vec{v}_T\mu\nu} \right)\;,
\end{align}
where $\delta \vec{v}_T = \delta \vec{v}/ \sqrt{\delta \vec{v}\cdot\delta \vec{v}}$, $D_{\vec{v}\mu}=(D_{\textrm{E}4}, \vec{v}(\vec{v}\cdot \vec{D}_{\textrm{E}}))$, and $ D_{\vec{v}_T \mu}=(0, \delta \vec{v}_T (\delta \vec{v}_T \cdot \vec{D}_{\textrm{E}}))$. 
The definition for the velocity-oriented field strength is understood from the definition of the covariant derivative. 
In the momentum scaling to the Fermi surface, $\vert D_{\vec{v}\mu} \vert \simeq \vert D_{\vec{v}_T\mu} \vert \ll \Lambda_{\textrm{QCD}}$ and the self-duality relationship of the velocity-oriented instanton field strength can be satisfied, which leads to the condensation of the surface zero-mode of each helicity basis.

\subsection{Scalar and vector channel in the instanton background}

Consider the interaction terms of HDET which lead to marginal interactions in the low-energy effective model. 
The simplest four-quark interaction can be written as follows:
 \begin{align}
 \mathcal{L}^{\textrm{E}}_{(\bar{\psi} \Gamma \psi)^2} & \simeq -\frac{1}{4\mu^2} \left[ \bar{\psi}_{+}(\vec{v})\gamma^{0} (i \, \slash \hspace{-0.25cm}  D_{  \vec{v}_T})^2 \psi_{+}(\vec{v})\right] \left[\bar{\psi}_{+}(-\vec{v})\gamma^{0} (i \,\slash \hspace{-0.25cm}  D_{  -\vec{v}_T})^2 \psi_{+}(-\vec{v})\right] \nonumber\\
 &=  \frac{1}{4\mu^2} \frac{1}{16} \left[  \bar{\psi}^{c}_{+}(-\vec{v}) \Gamma^{A} \psi^{a}_{+}(\vec{v}) \right] \left[  \bar{\psi}^{b}_{+}(\vec{v}) \Gamma^{B} \psi^{d}_{+}(-\vec{v}) \right] \left[ \Gamma^{A} \gamma^0   (i \, \slash \hspace{-0.25cm}  D_{  \vec{v}_T})^2_{ba}   \Gamma^{B} \gamma^0 (i \,\slash \hspace{-0.25cm}  D_{ -\vec{v}_T})^2_{cd}  \right]_{\alpha \alpha}\;,\label{4qintinst}
\end{align}
where the rectangular brackets denote the trace over the spinor ($\alpha$)  and color ($a$) indices. 
By choosing $\Gamma^{A}=\Gamma^{B}=1$ and color-singlet combinations, the scalar four-quark interaction can be expressed as
 \begin{align}
 \mathcal{L}^{\textrm{E}}_{(\bar{\psi} \psi)^2} & \Rightarrow - \frac{1}{4\mu^2} \frac{1}{16}  \left( \frac{1}{3} \right)^2  \left( \frac{g}{4} \right)^2[\gamma^0 \alpha^{k}  \gamma^0 \alpha^{n}]_{\alpha \alpha}[\tau^C \tau^D]_{aa} \epsilon^{ijk} \epsilon^{lmn}\nonumber\\
 &\qquad\quad \times\left[  \bar{\psi}_{+}(-\vec{v}) \psi_{+}(\vec{v}) \right] \left[  \bar{\psi}_{+}(\vec{v}) \psi_{+}(-\vec{v}) \right]   \left\langle \mathcal{G}^{C}_{\vec{v}_T ij} \mathcal{G}^{D}_{-\vec{v}_T lm} \right\rangle \nonumber\\
 &=-\frac{\pi}{3}\frac{1}{256}\frac{\alpha_s}{\mu^2} \left\langle {\mathcal{G}_{\vec{v}_T}}^2 \right\rangle \left[  \bar{\psi}_{+}(-\vec{v}) \psi_{+}(\vec{v}) \right] \left[  \bar{\psi}_{+}(\vec{v}) \psi_{+}(-\vec{v}) \right]\;,
 \label{pertcont}
\end{align}
where the following relations for the expectation value of the instanton field strength on both sides of the Fermi sphere are used:
\begin{align}
(i \,\slash \hspace{-0.25cm}  D_{ \vec{v}_T})^2&=D_{\vec{v}_T}^2-\frac{g}{2}\sigma_{\mu\nu}\mathcal{G}_{\vec{v}_T \mu\nu} = D_{\vec{v}_T}^2 +\frac{g}{4} \epsilon^{ijk} \alpha^k \mathcal{G}_{\vec{v}_T ij}\;,\\
 \left\langle \mathcal{G}^{C}_{\vec{v}_T ij} \mathcal{G}^{D}_{-\vec{v}_T lm} \right\rangle  &=- \frac{\delta^{CD}}{6} \left( \delta_{il} \delta_{jm} -v_{i} v_{l} \delta_{jm}  -v_{j} v_{m}\delta_{il}- \delta_{im} \delta_{jl}+ v_{i} v_{m} \delta_{jl}+ v_{j} v_{l} \delta_{im}  \right) \left\langle {\mathcal{G}_{\vec{v}_T}}^2 \right\rangle\;,
\end{align}
where $C,D=\{1,2,3\}$ denotes the adoint color index for the instanton field strength. 
In the perturbative regime ($\mu \gg \Lambda_{\textrm{QCD}} $), the strong coupling constant suppresses the contribution (\ref{pertcont}).
However,  at $\mu\simeq \mathcal{O}(\Lambda_{\textrm{QCD}})$, multi-instanton configurations~\cite{Diakonov:1983hh, Diakonov:1985eg, Shuryak:1989cx} and an increasing $\alpha_s$ in the low-momentum scaling to the Fermi surface can enhance the contribution  of this channel. 
This kind of interaction is of similar form as the effective interaction~\eqref{4qs}, which will be adopted as a scalar interaction in the simple model for the FRG analysis in Sec.~\ref{sec3}.

If  repulsive vector interactions near the Fermi surface  become dominant, the non-trivial scalar  channel could be suppressed in the RG flow to the IR limit.
Such a vector interaction can, for instance, be constructed from the vector current of the surface modes:
\begin{align}
 \bar{\psi} (x) \gamma^\mu \psi (x) \Longrightarrow &  \sum_{\vec{v}}   \bar{\psi} (\vec{v}, x) \gamma^\mu \psi (\vec{v}, x)\simeq   \bar{\psi}_{+} (\vec{v}, x) \gamma^{0} \psi_{+} (\vec{v}, x)\;,
\end{align}
where the relation $\vec{\alpha} \psi_{\pm} (\vec{v}, x) = \pm \vec{v} \psi_{\pm}  (\vec{v}, x)$ was used. 
For the zero-sound type interaction, we have two possible configurations for the Fermi velocities, (i) $\vec{v}_1+\vec{v}_2\simeq \vec{v}_3+\vec{v}_4 \simeq 0$, with $\vec{v}_1\simeq\vec{v}_3$ ($\vec{v}_2\simeq\vec{v}_4$), and
(ii) $\vec{v}_1\simeq\vec{v}_2\simeq \vec{v}_3\simeq\vec{v}_4$  (cf.\ caption of Fig.~\ref{f}(a)).
For case (i), the interaction Lagrangian can be constructed as
\begin{align}
\mathcal{L}_{(\bar{\psi} \gamma \psi)^2}&= \left[ \bar{\psi} (x) \gamma^\mu \psi (x) \right]\left[ \bar{\psi} (x) \gamma_\mu \psi (x) \right] \nonumber\\
&\Rightarrow  \sum_{\vec{v}} \frac{1}{\mu^2} \frac{g_v}{n_D}  \left[ \bar{\psi}_{+}  (\vec{v}, x) \gamma^\mu \psi_{+} (\vec{v}, x) \right] \left[ \bar{\psi}_{+} (-\vec{v}, x) \gamma_\mu \psi_{+} (-\vec{v}, x) \right]\nonumber\\
&\simeq \sum_{\vec{v}} \frac{ 2}{\mu^2} \frac{g_v}{n_D} \left[ \bar{\psi}_{+}  (\vec{v}, x) \gamma^{0} \psi_{+} (\vec{v}, x) \right] \left[ \bar{\psi}_{+} (-\vec{v}, x) \gamma^{0} \psi_{+} (-\vec{v}, x) \right]\;.
\end{align}
The QCD origin of this channel can be found not only from the instanton configuration but from the perturbative gluon exchange appearing in the following interaction:
 \begin{align}
\mathcal{L}^{\textrm{E}}_{(\psi^{\dagger} \psi)^2} & \simeq - \frac{1}{4\mu^2} \left[ \bar{\psi}_{+}(\vec{v})\gamma^{0} D_{\textrm{E}}^2 \psi_{+}(\vec{v})\right] \left[\bar{\psi}_{+}(-\vec{v})\gamma^{0}  D_{\textrm{E}}^2 \psi_{+}(-\vec{v})\right]\;.
\end{align}
The second case (ii), which corresponds to collinear forward scattering, actually does not appear at leading order since 
\begin{align}
\left[ \bar{\psi}_{+}  (\vec{v}, x) \gamma^\mu \psi_{+} (\vec{v}, x) \right]  \left[ \bar{\psi}_{+} (\vec{v}, x) \gamma^\mu \psi_{+} (\vec{v}, x) \right]\simeq V^2 \left[ \bar{\psi}_{+} (\vec{v}, x) \gamma^{0} \psi_{+}(\vec{v},x) \right]^2=0\;,
\end{align} 
but can appear  at  next-to-leading order. 
Because this interaction is constructed from the vector current, it becomes vanishingly small when the outer shell of the Fermi sphere becomes infinitesimally thin (Fig.~\ref{fig1}). 
One can guess the transverse vector interaction from Eq.~\eqref{4qintinst} but further analysis of such a kind of channel will be deferred to future work.
\newpage
\section{FRG analysis of effective model}~\label{sec3}

In this section, we construct a simple effective model by considering scalar- and vector-type four-quark contact interactions. 
The effective Lagrangian is bosonized via a Hubbard-Stratonovich transformation, and this effective form is taken as UV initial condition in the FRG analysis. 
First, we investigate the quark condensation amplitude in the mean-field approximation, assuming a homogeneous amplitude $\sigma$ in the FRG flow of the action. 
The case of an inhomogeneous amplitude is investigated by analyzing the scalar two-point function when allowing for spatial fluctuations. 
Finally, we investigate the case of a dynamical $\sigma$ by solving the FRG flow equation.

\subsection{Effective model Lagrangian and Hubbard-Stratonovich transformation}

For the general description, one may consider $\mu=\mu_q+\bar{\mu}$, $\mu_q \gg \bar{\mu}$ and write the effective Lagrangian in terms of the surface modes. 
The parameter $\bar{\mu}$ can be interpreted as the thickness of the momentum shell where one expects  quark condensation, cf.~Fig.~\ref{fig1}. 
Then, the effective  Lagrangian of the matter part can be expressed as follows~\cite{Schafer:2003jn}:
\begin{align}
\mathcal{L}_q &\simeq  \sum_{\vec{v}}\left[\bar{\psi}_{+}  (\vec{v}, x)  \gamma^0(i V^\mu D_\mu +\bar{\mu})\psi_{+}  (\vec{v}, x)  -\bar{\psi}_{+}  (\vec{v}, x)  \frac{\gamma^0}{2\mu_q+\bar{\mu}+i\bar{V}^\mu D_\mu}
\slash \hspace{-0.25cm}  D_\bot^2 \psi_{+}  (\vec{v}, x)  \right] \nonumber\\
&\qquad + \sum_{\vec{v}_{i}} \sum_{\Gamma \Gamma'}\frac{d^{\Gamma \Gamma'}(v_1,v_2,v_3,v_4)}{\mu_q^2}\left[\bar{\psi}_{+}  (\vec{v}_3, x) \Gamma \psi_{+}  (\vec{v}_1, x)  \right] \left[\bar{\psi}_{+}  (\vec{v}_4, x) \Gamma' \psi_{+}  (\vec{v}_2, x)  \right]+\ldots\;.
\label{eq:Lq}
\end{align}
From this point, we will only keep the effective four-quark interactions considered in the previous section and omit the subscript `$+$' on the positive quark modes. 
To describe the interaction between the quark and holes on the patches in opposite directions of the Fermi velocity, it is convenient to double the Fermi velocity sums and divide by 2.
Converting Eq.~\eqref{eq:Lq} into Euclidean space we then obtain
\begin{align}
\mathcal{L}_{\textrm{E}}=\frac{1}{2}\sum_{\vec{v}}& \bigg\{ \bar{\psi}(\vec{v}, x)  \left(-i {\slash \hspace{-0.2cm}  \partial}_{\textrm{E}}-\bar{\mu}\gamma^0 \right) \psi (\vec{v}, x)  + \bar{\psi}  (-\vec{v}, x)\left( -i {\slash \hspace{-0.2cm}  \partial}_{\textrm{E}} -\bar{\mu}\gamma^0 \right) \psi (-\vec{v}, x) \nonumber\\
&\quad- \frac{1}{\mu_q^2 } \frac{g_s}{ n_D} \left[ \bar{\psi}(-\vec{v}, x)   \psi(\vec{v}, x)   \right] \left[ \bar{\psi}(\vec{v}, x) \psi (-\vec{v}, x)  \right] - \frac{1}{\mu_q^2 } \frac{g_v}{n_D}  \left[ \bar{\psi}  (\vec{v}, x) \gamma_\mu \psi (\vec{v}, x) \right] \left[\bar{\psi}  (-\vec{v}, x) \gamma_\mu \psi (-\vec{v}, x) \right] \bigg\}\;.\label{modell}
\end{align}
Employing a Hubbard-Stratonovich transformation~\cite{1957SPhD....2..416S, Hubbard:1959ub} we obtain the bosonized form of this Lagrangian. Explicitly, we introduce auxiliary fields $\xi^s, \xi^v$ into the partition function of the model,
\begin{align}
\mathcal{Z}&=\textrm{const.}\int \mathcal{D}[ \xi^s, \xi^v, \bar{\psi}, \psi]\,  e^{-S_E}\, \exp\left(-\frac{\mu^2_q }{2g_s}\sum_{\vec{v}}\xi^{s}_{\vec{v}} \xi^{s}_{-\vec{v}} - \frac{\mu^2_q }{2g_v}\sum_{\vec{v}}\xi^{v}_{\vec{v}, \mu} \xi^{v}_{-\vec{v},\mu} \right) \;,
\end{align}
where $S_E \equiv \int d^4x_{\textrm{E}}\, \mathcal{L}_E$.
Then, we substitute $\xi^s_{\vec{v}}$ and $\xi^v_{\vec{v}}$ by the shifted auxiliary fields $\sigma^\pm$, $\omega_\mu^\pm$ via
\begin{align}
\xi^s_{\pm\vec{v}} & = h_s \sigma^{\pm}(x) +  \frac{1}{\mu^2_q } \frac{g_s}{\sqrt{n_D}}  \bar{\psi} (\mp\vec{v},x)  \psi (\pm\vec{v},x)\;,\label{hsts}\\
\xi^v_{\pm\vec{v},\mu} & = h_v \omega_{\mu}^{\pm}(x) +  \frac{1}{\mu^2_q } \frac{g_v}{\sqrt{n_D}}  \bar{\psi} (\pm\vec{v},x) \gamma_{\mu} \psi (\pm\vec{v},x)\;,\label{hstv}
\end{align}
to obtain the bosonized partition function
\begin{equation}
\mathcal{Z} =\textrm{const.} \int \mathcal{D}[ \sigma, \omega, \bar{\psi}, \psi] e^{-S_{b\textrm{E}}}\;,
\end{equation}
where $S_{b\textrm{E}}$ is the action corresponding to the bosonized Lagrangian
\begin{align}
\mathcal{L}_{b\textrm{E}}=\frac{1}{2}\sum_{\vec{v}}& \left[ \bar{\Psi}_{\vec{v}} \left (-i   \partial_{\mu} \Gamma_{\mu}-\bar{\mu}\Gamma^0+ \frac{h_s}{\sqrt{n_D}}\varphi + \frac{h_v}{\sqrt{n_D}}\omega_{\mu} \Gamma_{\mu} \right) \Psi_{\vec{v}} + \frac{\mu_q^2 h^2_s}{g_s}\sigma^{-}(x)\sigma^{+}(x)+ \frac{\mu_q^2 h^2_v}{g_v}\omega^{-}_{\mu}(x)\omega^{+}_{\mu}(x)\right]\;.\label{lb}
\end{align}
Here, we introduced the matrix representations
\begin{align}
 \Psi_{\vec{v}} & =
  \left( {\begin{array}{c}
  \psi_{+} (\vec{v},x)\\
  \psi_{+} (-\vec{v},x)\end{array}} \right),\quad \Gamma_{\mu}=\left( 
  {\begin{array}{cc}
0 & \gamma_{\mu} \\
  \gamma_{\mu} &  0 \\
  \end{array}} \right), \quad \varphi=\left( {\begin{array}{cc}
 \sigma^{-}(x)I & 0 \\
  0 &  \sigma^{+}(x)I \\
  \end{array} } \right), \quad \omega_{\mu}=\left( {\begin{array}{cc}
\omega^{+}_{\mu}(x)I & 0 \\
  0 &  \omega^{-}_{\mu}(x)I \\
  \end{array} } \right),
\end{align}
where $ \sigma^{-}(x) =( \sigma^{+}(x))^{\dagger}$ follows from Eq.~\eqref{hsts} and where $\omega^{\pm}_{\mu}(x)=\left(\omega^{\pm}_{4}(x),\vec{v} [\vec{v}\cdot \vec{\omega}^
{\pm}(x)] \right)$. 
One should note that the independent spin degrees of freedom have been reduced by a factor of 2, since half of the helicity space corresponds to the negative-energy modes, which have been integrated out. 
The quark mode along the opposite Fermi velocity, i.e., $  \psi({-\vec{v}},x)$, plays the role of the negative-energy modes of the vacuum theory.

\subsection{The exact RG flow equation in the mean-field approximation for the bosonic fields}

We first investigate the simplest configuration where the bosonic fields are approximated by constant mean fields: 
\begin{align}
 \varphi\Rightarrow \left( {\begin{array}{cc}
 \sigma I & 0 \\
  0 &  \sigma I \\
  \end{array} } \right)\;, \quad \omega_{\mu} \Rightarrow \left( {\begin{array}{cc}
\omega^{+}_{\mu}I & 0 \\
  0 &  \omega^{-}_{\mu}I \\
  \end{array} } \right)\;,
\end{align}
where $\sigma$ is a real-valued background field ($\sigma^{-}=(\sigma^{+})^{\dagger}=\sigma^{+}=\sigma$) and $\omega^{\pm}_{\mu}=\left(\omega_{4}, \pm \vec{v} \omega_{\parallel} \right)$ with  $\omega^{\pm}_{4}=\omega_4$ and $\omega_{\parallel}\equiv \vec{v}\cdot \vec{\omega}^{+}=- \vec{v}\cdot \vec{\omega}^{-}$. 
Then, to leading order the effective average action  corresponding to the bosonized Lagrangian density~\eqref{lb} reads
\begin{align}
\Gamma^{m}_k = \int d^4x_{\textrm{E}} \bigg[ \frac{1}{2}\sum_{\vec{v}} \bar{\Psi}_{\vec{v}}\left (-i  \Gamma_{\mu} \partial_{\mu} -\bar{\mu}\Gamma^0+ \frac{h_s}{\sqrt{n_D}}\varphi + \frac{h_v}{\sqrt{n_D}}\omega_{\mu} \Gamma_{\mu} \right) \Psi_{\vec{v}} + 2 \pi \mu_q^2 \frac{h^2_v}{g_v}(\omega_4^2-\omega_{\parallel}^2)
+U_\sigma(k,\sigma)\bigg]\;,\label{Ab}
\end{align}
where
$U_\sigma(k,\sigma)$ is the RG time-dependent potential for the $\sigma$-field, which has the UV initial condition
\begin{equation}
U_\sigma(k=\mu_q, \sigma) = 2 \pi \mu_q^2
\frac{h_s^2}{g_s} \sigma^2\;.
\end{equation}
The coarse graining of $\Gamma^{m}_k$ with respect to the RG time can be described by including the regulator term:
\begin{align}
\Gamma^{f}_{k}&=\Gamma^{m}_k+ \Delta \Gamma^{f}_{k},\\
\Delta \Gamma^{f}_{k} &=  \frac{1}{2}\sum_{\vec{v}} \int \frac{d^4p_{\textrm{E}}}{(2\pi)^4} \bar{\Psi}_{\vec{v}}(-p_{\parallel}) R^{f}_{\vec{v}}(k,p) \Psi_{\vec{v}}(p_{\parallel}). 
\end{align}
The corresponding exact RG flow equation~\cite{Wetterich:1991be, Wetterich:1992yh, Ellwanger:1993mw} is
\begin{align}
\partial_t \Gamma^{f}_{k}  = - \textrm{STr} \left\{ \left[\partial_t R^{f}_{\vec{v}}(k,p) \right] \left[ \Gamma^{m (2)}_{\bar{\Psi}\Psi}(k, \vec{v})+R^{f}_{\vec{v}}(k,p) \right]^{-1} \right\}\;,\label{eff}
\end{align}
where $\Gamma^{m (2)}_{\bar{\Psi}\Psi}(k, \vec{v})$ denotes the second  functional derivative of $\Gamma^{m}_{k}$ with respect to the fermionic fields evaluated at $\bar{\Psi}_{\vec{v}}=\Psi_{\vec{v}}=0$. `STr' denotes the supertrace including the summation over the spin, color, and flavor indices and the integration over $p_{\parallel}$ and $\vec{v}$.
The fermionic regulator $ R^{f}_{\vec{v}}(k,p)$ can be derived from the well-known Litim regulator~\cite{Litim:2000ci, Litim:2001up} as 
\begin{align}
R^{f}_{\vec{v}}(k,p) = \frac{ \beta\delta_{n'n} \delta^{(2)}(\vec{v}^{\, \prime}-\vec{v}) \, 2\pi \delta(p_{\parallel}' - p_{\parallel})}{2} \left( {\begin{array}{cc}
0 & i p_{\parallel}\gamma_{4} \\
 -i p_{\parallel}\gamma_{4}&  0 \\
  \end{array} } \right) r_f(k, p_{\parallel})\;,
\end{align}
where $p_{\parallel}=\vec{v}\cdot \vec{p}$, $k=\Lambda e^{-t}$, and $ r_f(k,p_{\parallel})$ is given as
\begin{align}
1+r_f(k,p_{\parallel})&=\sqrt{1+r_b(k,p_{\parallel})}\;,\\
r_b(k,p_{\parallel})&\equiv \left( \frac{k^2}{p_{\parallel}^2}-1 \right) \Theta\left(\frac{k^2}{p_{\parallel}^2}-1 \right)\;,
\end{align}
and the integration measure contained in the supertrace in Eq.\ (\ref{eff}) is understood as
\begin{align}
\int \frac{d^4p_{\textrm{E}}}{(2\pi)^4}\beta \delta_{n'n}(2\pi)^3\delta^{(3)}(\vec{p}^{\, \prime}-\vec{p}) & \equiv \frac{1}{\beta} \sum_{n} \sum_{\vec{v}}\int_{-\mu_q}^{\mu_q}\frac{d p_{\parallel}}{2\pi} \beta \delta_{n'n} \delta^{(2)} (\vec{v}^{\, \prime}-\vec{v}) \, 2\pi\delta (p_{\parallel}'-p_{\parallel})\;,
\end{align}
where we used $\delta^{(2)}(\vec{p}_{\bot})=(1/\mu_q^2)\delta^{(2)} (\vec{v})$ and $d^2p_{\bot} = \mu_q^2 d \vec{v}$. 
Note the following useful relations:
\begin{align}
2\partial_t r_f(k,p_{\parallel}) \left[1+r_f(k,p_{\parallel})\right]&= \partial_t r_b(k,p_{\parallel}) =-  \frac{2 k^2}{p_{\parallel}^2}   \left[  \Theta\left(\frac{k^2}{p_{\parallel}^2}-1 \right)+\left(\frac{k^2}{p_{\parallel}^2}-1 \right)\delta\left(\frac{k^2}{p_{\parallel}^2}-1 \right) \right]\;,\\
\partial_t r_f(k,p_{\parallel})&= \frac{\partial_t r_b(k,p_{\parallel})}{2\sqrt{1+r_b(k,p_{\parallel})}} =\bigg\{ {\begin{array}{cc}
0 & \vert p_{\parallel} \vert > k\\
-\frac{k}{p_{\parallel}}  &  \vert p_{\parallel} \vert \leq k \\
  \end{array} }\;.
\end{align}
$\Gamma^{m (2)}_{\bar{\Psi}\Psi}(k, \vec{v})+R^{f}_{\vec{v}}(k,p)$ can be expressed explicitly as follows:
\begin{align}
\Gamma^{m (2)}_{\bar{\Psi}\Psi}(k, \vec{v})& +R^{f}_{\vec{v}}(k,p)\nonumber\\
& =  \frac{ \beta\delta_{n'n} \delta^{(2)}(\vec{v}^{\, \prime}-\vec{v})\, 2\pi \delta(p_{\parallel}' - p_{\parallel}) }{2} \Bigg[ \left( {\begin{array}{cc}
0 & \gamma_{4} \left[p_4+i\bar{\mu}+ip_{\parallel}(1+r_f(k, p_{\parallel}) \right] \\
  \gamma_{4} \left[p_4+i\bar{\mu}-ip_{\parallel}(1+r_f(k, p_{\parallel}) \right]   &  0 \\
  \end{array} } \right) \nonumber\\
  &\qquad\qquad\qquad\qquad\qquad\qquad\qquad\qquad+ \frac{h_s}{\sqrt{n_D}} \left( {\begin{array}{cc}
\sigma  I & 0 \\
  0 & \sigma  I \\
  \end{array} } \right) + \frac{h_v}{\sqrt{n_D}} \left( {\begin{array}{cc}
0 & \gamma_{4} \left(\omega_4+i\omega_{\parallel} \right) \\
 \gamma_{4} \left(\omega_4+i\omega_{\parallel} \right)  &  0 \\
  \end{array} } \right)    \Bigg]\;,
\end{align}
and the inverse can be obtained as
\begin{align}
&\left[ \Gamma^{m (2)}_{\bar{\Psi}\Psi}(k, \vec{v})+R^{f}_{\vec{v}}(k,p) \right]^{-1} \nonumber\\
&\qquad\qquad=  \frac{2 \beta\delta_{n'n} \delta^{(2)}(\vec{v}^{\, \prime}-\vec{v}) \,2\pi \delta(p_{\parallel}' - p_{\parallel}) }{{\bar{p}_4}^2+{\bar{p}_{\parallel}}^2+ \frac{h^2_s}{n_D}\sigma^{2}}   \left[ \frac{h_s}{\sqrt{n_D}} \left( {\begin{array}{cc}
\sigma  I & 0 \\
  0 & \sigma  I \\
  \end{array} } \right)-\left( {\begin{array}{cc}
0 & \gamma_{4} \left(\bar{p}_4  +i\bar{p}_{\parallel}\right) \\
 \gamma_{4} \left(\bar{p}_4 -i\bar{p}_{\parallel}\right)   &  0 \\
  \end{array} } \right) \right]\;,
\end{align}
where $\bar{p}_4$ and $\bar{p}_{\parallel}$ denote
\begin{align}
  \bar{p}_4 &=p_4+i\left[\bar{\mu}+ \frac{h_v}{\sqrt{n_D}} \left(\omega_0+\omega_{\parallel} \right) \right]\;,\\
  \bar{p}_{\parallel}&=p_{\parallel}\left[1+r_f(k, p_{\parallel})\right]\;.
\end{align}
Then, after taking the trace on the other indices (considering an additional factor 1/2), Eq.~\eqref{eff} can be expressed as
\begin{align}
\partial_t \Gamma^{f}_{k}  &=- \beta (2\pi)^3 \delta^{(3)}(0) \frac{1}{\beta} \sum_n \frac{\mu_q^2}{(2\pi)^2}\sum_{\vec{v}} \int_{-\mu_q}^{\mu_q}\frac{d p_{\parallel}}{2\pi} \frac{4 n_D   p_{\parallel}^2[1+r_f(k, p_{\parallel})]  \partial_t  r_f(k, p_{\parallel})}{\left\{p_4+i\left[\bar{\mu}+ \frac{h_v}{\sqrt{n_D}}(\omega_0+\omega_{\parallel} )\right]\right\}^2+\left[p_{\parallel}(1+r_f(k, p_{\parallel}))\right]^2+ \frac{h^2_s}{n_D}\sigma^{2}}\nonumber\\
& =  \beta (2\pi)^3 \delta^{(3)}(0)  \frac{2n_D \mu_q^2}{\pi^2} \frac{ k(t)^3  }{E_k(\sigma)}\left\{1-n_{\textrm{f}}\left[\beta \left(E_k(\sigma)+\tilde{\mu}(\omega_{\parallel\mu})\right) \right] - n_{\textrm{f}}\left[\beta \left(E_k(\sigma)-\tilde{\mu}(\omega_{\parallel\mu})\right) \right] \right\}\;,\label{fleq}
\end{align}
where $n_{\textrm{f}}[x]=1/(e^x+1)$, $E_k(\sigma)=\left[ k(t) ^2+ \frac{h^2_s}{n_D}\sigma^{2} \right]^{\frac{1}{2}}$, and $(2\pi)^3\delta^{(3)}(0) $ denotes the system's volume. 
In mean-field approximation, $\frac{h_v}{\sqrt{n_D}} \left(\omega_0+\omega_{\parallel} \right)$  can be absorbed into the shell parameter $\bar{\mu}$, as there is no kinetic fluctuation of the bosonic fields: $\tilde{\mu}( \omega_{\parallel\mu}) = \bar{\mu} + \frac{h_v}{\sqrt{n_D}} \left(\omega_0+\omega_{\parallel} \right)$. 
We will assume the condition $\tilde{\mu}( \omega_{\parallel\mu}) \rightarrow 0$ in the limit  $\bar{\mu}\rightarrow 0$ to make the medium contribution from the momentum shell vanish when the parameter $\bar{\mu} \rightarrow 0$.

\subsection{Analytic investigation and numerical results in the mean-field approximation for the bosonic fields}

As the quantum fluctuation from the bosonic part is missing in the mean-field approximation, if one neglects the wave function renormalization of the fermionic fields, the RG-time dependence of the effective action comes only from the bosonic potential term: 
\begin{align}
\partial_{t} \Gamma^{f}_{k}   &=   \beta (2\pi)^3 \delta^{(3)}(0) \partial_{t} U( k(t), \sigma, \omega_{\parallel\mu})\;,
\end{align}
with the UV initial condition
\begin{align}
U( k=\mu_q,\sigma, \omega_{\parallel\mu}) &=  2\pi \mu_q^2\left[ \frac{ h^2_s}{g_s}\sigma^{2}+ \frac{ h^2_v}{g_v}\left(-\omega_0^2-\omega_{\parallel}^2 \right)\right]\;.
\end{align} 
Then, the flow equation~\eqref{fleq} can be integrated to give the following result:
\begin{align}
U( 0,\sigma, \omega_{\parallel\mu}) &= U( \mu_q,\sigma, \omega_{\parallel\mu}) +  \frac{ 2n_D\mu_q^2}{\pi^2}\int^{ \mu_q}_{0}dk \frac{ k^2}{E_{k}(\sigma)}\bigg\{1-n_{\textrm{f}}\left[\beta \left(E_{k}(\sigma)+\tilde{\mu}(\omega_{\parallel\mu})\right) \right] - n_{\textrm{f}}\left[\beta \left(E_{k}(\sigma)-\tilde{\mu}(\omega_{\parallel\mu})\right) \right] \bigg\}\;,\label{irpot}
\end{align}
where $k(t)$ flows from $k(0)= \mu_q$ to $k(\infty)= 0$. 
Since we are considering the situation where $\bar{\mu} \ll \mu_q$ and $T \ll \mu_q$, the contribution from the momentum shell around the quark Fermi sphere ($\bar{\mu} > 0$) will be treated as the ``medium correction'' to the thin-shell limit ($T\rightarrow 0,~\bar{\mu} \rightarrow 0$) of the potential (see Fig.~\ref{fig1}).
The IR potential in the thin-shell limit can be identified as follows:
\begin{align}
U_{\textrm{TS}}( 0,\sigma, \omega_{\parallel\mu}) &= 2\pi \mu_q^2\left[ \frac{ h^2_s}{g_s( \mu_q)}\sigma^{2}+ \frac{h^2_v}{g_v(  \mu_q)}\left(-\omega_0^2-\omega_{\parallel}^2\right)\right]+\frac{2n_D\mu_q^2}{\pi^2} \left[ \mu_q \left(\mu_q^2+\frac{h_s^2}{n_D}\sigma^2 \right)^{\frac{1}{2}}-\int_{0}^{\mu_q}dk  \left(k^2+\frac{h_s^2}{n_D}\sigma^2 \right)^{\frac{1}{2}} \right]\;.
\end{align}
We investigate the scalar potential first. 
If one assumes a nontrivial minimum  $\sigma_{0}\neq 0$,  the  corresponding $2\pi/g_s( \mu_q)$  can be found from the  constraint $\partial_{\sigma} U_{\textrm{TS}} \vert_{\sigma=\sigma_{0}}=0$:
\begin{align}
\frac{2\pi}{g_s( \mu_q)}\bigg\vert_{\textrm{TS}}&=-\frac{1}{\pi^2} \left\{  \left(1+\frac{h_s^2}{\mu_q^2}\frac{\sigma_{0}^2 }{n_D} \right)^{-\frac{1}{2}}-\ln \left[ \frac{\frac{h_s}{\mu_q}\frac{\sigma_{0} }{\sqrt{n_D}} }{\left(1+\frac{h_s^2}{\mu_q^2}\frac{\sigma_{0}^2 }{n_D} \right)^{\frac{1}{2}}-1} \right]\right\}\;.\label{uvgs}
\end{align}
Keeping $h_s$ fixed and taking the large-$\mu_q$ limit, the following form can be found:
\begin{align}
\frac{2\pi}{g_s( \mu_q)}\bigg\vert_{\textrm{TS}}&=-\frac{1}{\pi^2} \left[ 1-\ln \left( \frac{2\sqrt{n_D}}{\sigma_{0}}\frac{\mu_q  }{  h_s }\right) \right] + \mathcal{O}\left( \frac{h_s^2}{\mu_q^2} \right)+\ldots\;,
\end{align}
where $g_s( \mu_q) \rightarrow 0$ in the limit of $\mu_q / h_s\rightarrow \infty$ as discussed in Refs.~\cite{Stoll:2021ori, Wolff:1985av}. 
A possibly appearing divergence of $U_{\textrm{TS}}( 0,\sigma, \omega_{\parallel\mu})$ in the limit $\mu_q / h_s\rightarrow \infty$ is regularized by Eq.~\eqref{uvgs}.

\begin{figure}
\includegraphics[height=5.6cm]{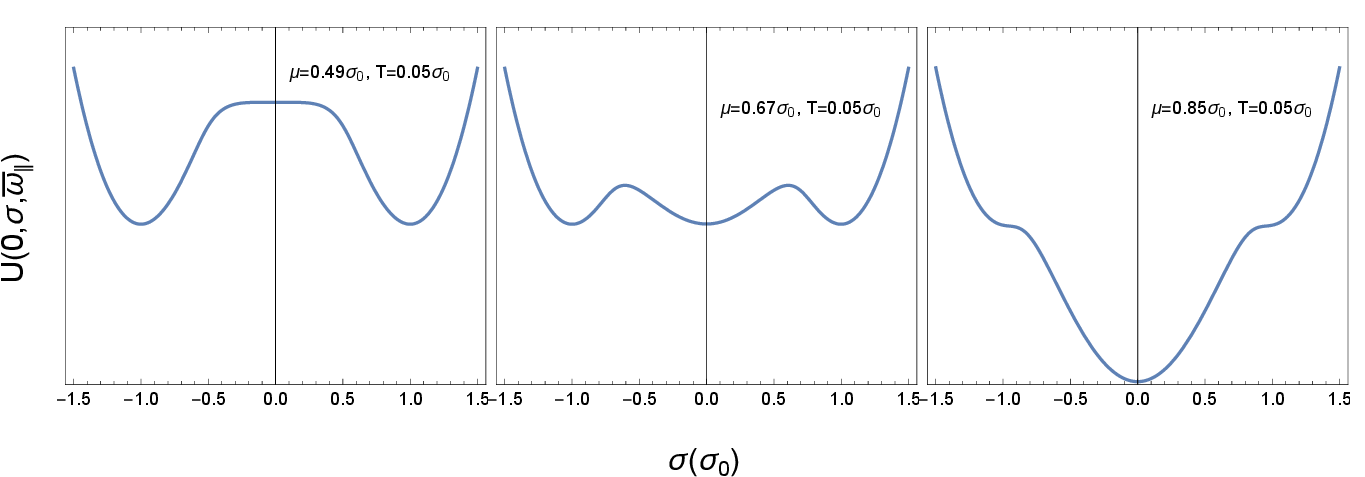}
\caption{(Color online) $U( 0,\sigma, \bar{\omega}_{\parallel\mu})$  for $\mu=400~\textrm{MeV}$ and $T=0.05\sigma_{0}$. 
From left to right, $\bar{\mu}=0.65\sigma_{0}$, $\bar{\mu}=0.67\sigma_{0}$, and $\bar{\mu}=0.85 \sigma_{0}$. 
In this low-$T$ regime, the system develops a local minimum at $\sigma=0$ (left spinodal point at $\bar{\mu}=0.65\sigma_{0}$), which becomes the global minimum at the transition (first-order phase transition). Around $\bar{\mu}=0.85 \sigma_{0}$, the local minima (which were the global minima before the transition) disappear and only the global minimum remains (right spinodal point).}\label{fig3}
\end{figure}

Since a nontrivial $\bar{\omega}_{\parallel\mu}$-dependence only comes from the shell contribution, in the thin-shell limit ($T\rightarrow 0,~\bar{\mu} \rightarrow 0$) $\bar{\omega}_{\parallel\mu}=(\bar{\omega}_0, \bar{\omega}_{ \parallel}) \Rightarrow (0,0)$ and the corresponding $2\pi/g_v( \mu_q) \rightarrow 0$ . 
Considering the shell contribution, one can assume $\bar{\omega}_{\parallel\mu}=(\bar{\omega}_0, \bar{\omega}_{ \parallel})\neq(0,0)$ and find the corresponding $2\pi/g_v( \mu_q)\neq0$ from the  constraints $\partial_{\omega_{0}} U_{\textrm{shell}} \vert_{\omega_0=\bar{\omega}_{0}}=0$ and  $\partial_{\omega_{\parallel}} U_{\textrm{shell}} \vert_{\omega_{\parallel}=\bar{\omega}_{\parallel}}=0$:
\begin{align}
\frac{2\pi}{g_v( \mu_q)}&=\frac{1}{\pi^2} \frac{\sqrt{n_D}}{h_v \bar{\omega}_{\parallel}} \Bigg( \mu_q \left\{ n_{\textrm{f}}\left[\beta \left(E_{ \mu_q}(\sigma)-\tilde{\mu}(\bar{\omega}_{\parallel\mu})\right)\right] -n_{\textrm{f}}\left[\beta \left(E_{ \mu_q}(\sigma)+\tilde{\mu}(\bar{\omega}_{\parallel\mu})\right) \right] \right\}\nonumber\\
&\qquad\qquad\qquad-\int_0^{\mu_q}dk \left\{ n_{\textrm{f}}\left[\beta \left(E_k(\sigma)-\tilde{\mu}(\bar{\omega}_{\parallel\mu})\right)\right]-n_{\textrm{f}}\left[\beta \left(E_k(\sigma)+\tilde{\mu}(\bar{\omega}_{\parallel\mu})\right\} \right] \right) \Bigg)\;,\label{cpv}
\end{align}
from which one can check that $2\pi/g_v( \mu_q) \leq 0$ for $\bar{\omega}_{\parallel}\geq 0$ and the asymptotic behavior $2\pi/g_v( \mu_q) \rightarrow0$ in the thin-shell limit ($T\rightarrow 0,~\bar{\mu} \rightarrow 0,~ \bar{\omega}_{\parallel \mu}\rightarrow0$), which guarantees the repulsive nature of the $\omega^2$-potential in $U(\mu_q ,\sigma, \omega_{\parallel\mu}) $ and implies the absence of an $\omega^2$-potential in the thin-shell limit, respectively. 
The thin-shell limit of the IR potential can be summarized as follows:
\begin{align}
U_{\textrm{TS}}( 0,\sigma, \omega_{\parallel\mu}) &=  
-\frac{\mu_q^2 (h_s \sigma)^2}{\pi^2} \left\{ \left(1+\frac{h_s^2}{\mu_q^2}\frac{\sigma_{0}^2 }{n_D} \right)^{-\frac{1}{2}}-\ln \left[ \frac{\frac{h_s}{\mu_q}\frac{\sigma_{0} }{\sqrt{n_D}} }{\left(1+\frac{h_s^2}{\mu_q^2}\frac{\sigma_{0}^2 }{n_D} \right)^{\frac{1}{2}}-1} \right] + \ln \left[ \frac{\frac{h_s}{\mu_q}\frac{\sigma }{\sqrt{n_D}} }{\left(1+\frac{h_s^2}{\mu_q^2}\frac{\sigma^2 }{n_D} \right)^{\frac{1}{2}}-1} \right]\right\}
\nonumber\\
&\qquad+\frac{\mu_q^4 n_D}{\pi^2}  \left(1+\frac{h_s^2}{\mu_q^2}\frac{\sigma^2 }{n_D} \right)^{\frac{1}{2}},
\end{align}
which, from a technical point of view, corresponds to the IR vacuum potential discussed in Refs.~\cite{Stoll:2021ori, Wolff:1985av}. 
The full IR potential~\eqref{irpot} can be evaluated as follows:
\begin{align}
U( 0,\sigma, \omega_{\parallel\mu}) &= U_{\textrm{TS}}( 0,\sigma, \omega_{\parallel\mu}) \nonumber\\
&\quad-\frac{\mu_q^2 (h_s \sigma)^2}{\pi^2}\int_{0}^{\mu_q} dk \Bigg( \frac{ k^2}{E_{k}(\sigma_{0})^3} \left\{n_{\textrm{f}}\left[\beta \left(E_k(\sigma_{0})+\tilde{\mu}(\omega_{\parallel\mu})\right) \right]+n_{\textrm{f}}\left[\beta \left(E_k(\sigma_{0})-\tilde{\mu}(\omega_{\parallel\mu}) \right) \right] \right\}\nonumber\\
&\qquad\qquad\qquad\qquad \qquad+  \frac{\beta k^2 }{E_{k}(\sigma_{0})^2} \left\{n_{\textrm{f}}\left[\beta \left(E_k(\sigma_{0})+\tilde{\mu}(\omega_{\parallel\mu}) \right) \right]^2+n_{\textrm{f}}\left[\beta \left(E_k(\sigma_{0})-\tilde{\mu}(\omega_{\parallel\mu})\right) \right]^2 \right\} \Bigg)\nonumber\\
&\quad+\frac{\mu_q^2}{\pi^2} \frac{h_v\sqrt{ n_D}}{\bar{\omega}_{ \parallel}} \left(-\omega_0^2-\omega_{\parallel}^2 \right) \Bigg( \mu_q \left\{ n_{\textrm{f}}\left[\beta \left(E_{ \mu_q}(\sigma)-\tilde{\mu}(\bar{\omega}_{\parallel\mu})\right) \right] -n_{\textrm{f}}\left[\beta \left(E_{ \mu_q}(\sigma)+\tilde{\mu}(\bar{\omega}_{\parallel\mu})\right) \right] \right\}\nonumber\\
&\qquad\qquad\qquad\qquad\qquad\qquad\qquad -\int_0^{\mu_q}dk \left\{ n_{\textrm{f}}\left[\beta \left(E_k(\sigma)-\tilde{\mu}(\bar{\omega}_{\parallel\mu})\right)\right]-n_{\textrm{f}}\left[\beta \left(E_k(\sigma)+\tilde{\mu}(\bar{\omega}_{\parallel\mu})\right) \right] \right\} \Bigg)\nonumber\\
&\quad-\frac{2n_D\mu_q^2}{\pi^2} \int_{0}^{\mu_q} dk \frac{ k^2 }{E_{k}(\sigma)} \left\{n_{\textrm{f}}\left[\beta \left(E_k(\sigma)+\tilde{\mu}(\omega_{\parallel\mu})\right)\right] +  n_{\textrm{f}}\left[\beta \left(E_k(\sigma)-\tilde{\mu}(\omega_{\parallel\mu})\right)\right] \right\}\;.\label{mfpot}
\end{align}

\begin{figure}
\includegraphics[height=5.6cm]{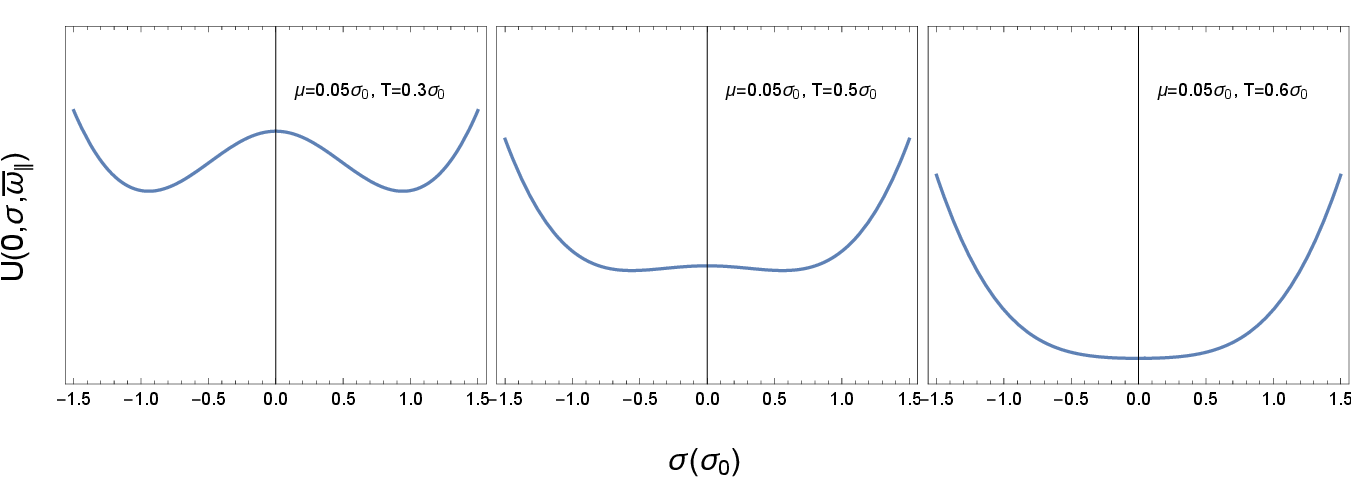} 
\caption{(Color online) $U( 0,\sigma, \bar{\omega}_{\parallel\mu})$  for $\mu=400~\textrm{MeV}$ and $\bar{\mu}=0.05\sigma_{0}$. 
From left to right, $T=0.3\sigma_{0}$, $T=0.5\sigma_{0}$, and $T=0.6\sigma_{0}$. 
In this low-$\bar{\mu}$ regime, the system does not develop another local minimum and the nontrivial global minima merge at the second-order phase transition.}\label{fig4}
\end{figure}

To investigate the nontrivial minima of $U( 0,\sigma, \omega_{\parallel\mu})$, we fix some of the parameters as follows: $h_s=\sqrt{6},~h_v=\sqrt{6},~n_D=6$. 
(Since $h_s$ and $h_v$ can be absorbed into the bosonic fields, these Yukawa parameters can be adjusted freely.) 
As one can see from the potential~\eqref{mfpot}, the qualitative behavior of the contribution from the momentum shell is determined by the relative size of $\bar{\mu}$ and $T$ compared to the nontrivial minimum $\sigma_0$. 
Here we fix the scale $\sigma_0 \ll \mu_q$ and express $\bar{\mu}$ and $T$ in units of $\sigma_0$. 
Considering that $ \langle \psi^{\dagger}(\vec{v}) \psi(\vec{v}) \rangle$ corresponds to the quark number density of the outer shell, the scale of $\bar{\omega}_0$ can be estimated from Eq.~\eqref{hstv}:
\begin{align}
\bar{\omega}_0\simeq & \frac{1}{\mu_q^2}\frac{g_v}{\sqrt{n_D}}  \langle \psi^{\dagger}(\vec{v}) \psi(\vec{v})  \rangle=\frac{g_v \sqrt{n_D}}{\pi^2}\bar{\mu}\;,
\end{align} 
where $\langle \psi^{\dagger}(\vec{v}) \psi(\vec{v})  \rangle= (n_D/\pi^2)\mu_q^2 \bar{\mu}$ in the massless limit. 
As $ \langle\bar{\psi} (\vec{v}) \gamma^{\mu} \psi(\vec{v}) \rangle =V^{\mu} \langle \psi^{\dagger}(\vec{v}) \psi(\vec{v})  \rangle$, in the mean-field approximation we will assume $\bar{\omega}_{\parallel}=\bar{\omega}_0$. 
We will ignore the RG-time dependence of $g_v$ and set $g_v \sqrt{n_D}=0.3$ for the sake of simplicity.

We first investigate the two different limits ($T\rightarrow 0, \bar{\mu} \neq0$) and ($T\neq0, \bar{\mu}\rightarrow 0 $). 
As one can observe in Fig.~\ref{fig3}, the system undergoes a first-order phase transition around ($T\rightarrow 0, \bar{\mu} =0.65\sigma_{0}$), where the nontrivial global minimum $\bar{\sigma}$ vanishes. 
In this direction, the system develops a local minimum at $\sigma=0$, which becomes the global minimum at the phase boundary. 
The two local minima which were the nontrivial global minima before the phase transition persist until around $ \bar{\mu} =0.85\sigma_{0}$ (right spinodal point). 

\begin{figure}
\includegraphics[height=6cm]{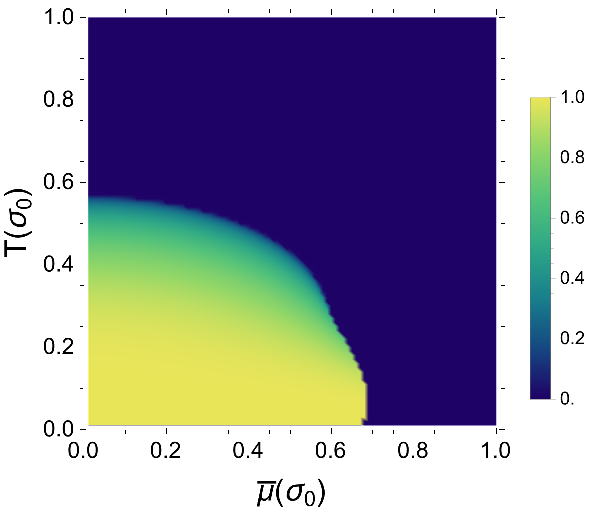} 
\includegraphics[height=6cm]{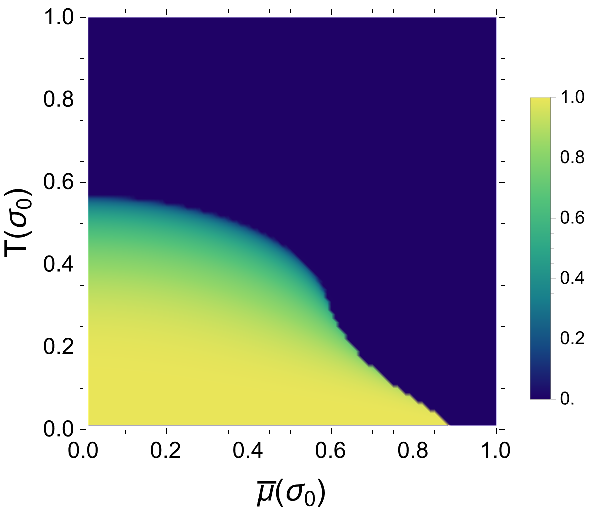}  
\caption{(Color online) Contour plot of the magnitude of the condensate $\bar{\sigma}$ at the local or global minima in the ($\bar{\mu},T$)-plane  for $\mu=400~\textrm{MeV}$. 
Left: the magnitude of $\bar{\sigma}$ at the nontrivial global minimum, which shows a first-order phase boundary in the low-$T$ regime. 
Right: the magnitude of $\bar{\sigma}$ at the nontrivial local minimum corresponding to the right spinodal boundary, which agrees with the global minimum to the left of the tricritical point ($T=0.25\sigma_{0}$, $\bar{\mu}=0.63\sigma_{0}$). As the shell thickness $\bar{\mu}$  increases in the low-$T$ regime, the system undergoes a first-order phase transition (Fig~\ref{fig3}). The local minimum corresponding to the right spinodal exists until around $\bar{\mu}\simeq 0.85 \sigma_0$.}\label{fig5}
\end{figure}

In the other limit ($T\neq0, \bar{\mu}\rightarrow 0 $) a second order phase transition occurs around ($T\simeq0.6\sigma_{0}, \bar{\mu} \rightarrow 0$), cf.\ Fig.~\ref{fig4}.
In this direction, the system does not develop any additional local minimum and the nontrivial minima disappear simultaneously at the phase boundary. This qualitative behavior is summarized in the phase diagrams shown in Fig.~\ref{fig5}. 

On the left-hand side of Fig.~\ref{fig5} we show the magnitude of $\bar{\sigma}$ at the nontrivial global minimum.
On the right-hand side, we show the magnitude of $\bar{\sigma}$ at the nontrivial local minimum corresponding to the right spinodal boundary, which agrees with the global minimum to the 
left of a (tri-)critical point ($T=0.25\sigma_{0}$, $\bar{\mu}=0.63\sigma_{0}$). 
This means there is a second-order phase transition in the high-$T$, low-$\bar{\mu}$ region and a first-order phase transition in the low-$T$, high-$\bar{\mu}$ region, separated by the tricritical point at ($T=0.25\sigma_{0}$, $\bar{\mu}=0.63\sigma_{0}$). 
The phase diagram is very similar to the standard one of the GN-model in mean-field approximation and assuming a spatially homogeneous condensate, except that the role of the ordinary chemical potential $\mu_q$ is now assumed by the shell-thickness parameter $\bar{\mu}$.

\subsection{Analysis of the bosonic two-point function for investigation of an inhomogeneous $\sigma(x)$ configuration}

\begin{figure}
\includegraphics[height=5.0cm]{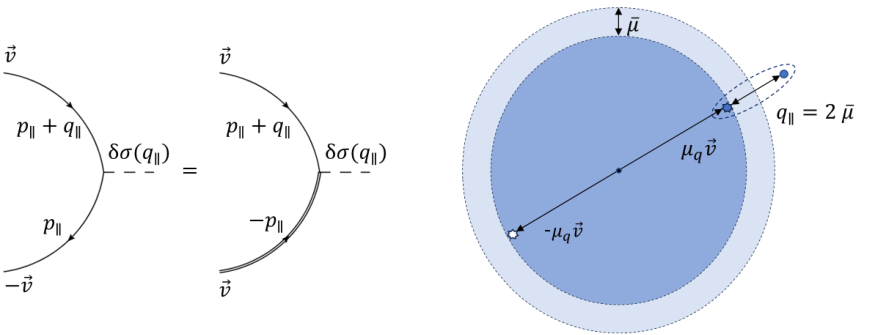} 
\caption{(Color online) Left: Diagrammatic representation of the $\delta\sigma(q_{\parallel})$ fluctuation. 
Each momentum is defined with respect to the Fermi surface ($\vert \vec{p}_{F}\vert = \mu_q$). 
A single line denotes the quark propagation and a double line denotes the hole propagation along the $\vec{v}$ direction.  Right: Illustration of a possible configuration of the quark ($ \psi_{+} (\vec{v}, q_{\parallel}=2\bar{\mu})$) and the hole ($ \bar{\psi}_{+} (-\vec{v}, q_{\parallel}\simeq 0)$) states forming $\delta\sigma(q_{\parallel})$ near the Fermi surface ($\bar{\mu} \ge 0.7\sigma_{0}$, $T \le 0.22\sigma_{0}$). 
The empty star near the Fermi surface at $ -\mu_q \vec{v}$ denotes the hole and the (dark blue) filled star at $\mu_q \vec{v}$  denotes its realization as an anti-particle state with respect to the ground state. 
The (dark blue) filled circle denotes the excited quark state. 
Since the hole state should lie inside the Fermi sphere, the anti-particle state can carry at most a momentum $\mu_q \vec{v}$ ($q_{\parallel}=0$). 
The excited quark state can carry 
a momentum up to $2\mu_q \vec{v}$ ($q_{\parallel}=\mu_q$) in this effective model, but when the momentum difference between the anti-particle and quark becomes $ q_{\parallel}=2\bar{\mu}$, the oscillating mode with the dominant wave vector $ 2\bar{\mu} \vec{v}$ would be an even more stable configuration. 
}\label{fig6}
\end{figure}

The next step is to consider a spatially inhomogeneous fluctuation $\delta \sigma(x)$ around a homogeneous mean-background field $\sigma$~\cite{Schnetz:2004vr,Nickel:2009ke, Nickel:2009wj, Buballa:2014tba, Buballa:2018hux, Koenigstein:2021llr}:
\begin{align}
\varphi\Rightarrow\left( {\begin{array}{cc}
 [\sigma+\delta\sigma(x)]I & 0 \\
  0 &  [\sigma+\delta\sigma(x) ]I \\
  \end{array} } \right)\;,
\end{align}
where the fluctation $\delta\sigma(x)$ can be related to the  momentum difference between the quark and hole modes in the outer shell of the Fermi sphere (cf.\ Figs.~\ref{fig1} and~\ref{fig6}). 
If the coefficient of the term quadratic in $\delta\sigma(x)$ in the effective average action, evaluated on the homogeneous background $\sigma$, becomes negative in some region of momentum $q_{\parallel}$, the homogeneous ground state is unstable and there is a more stable inhomogeneous configuration of the field $\sigma(x)$. 
The coefficient of the second-order term in the $\delta\sigma(x)$ fluctuation is the bosonic two-point function 
\begin{align}
\Gamma^{m (2)}_{(\delta\sigma)^2}(q_{\parallel})=\beta (2\pi)^3\delta^{(3)}(0)\left\{ \frac{ 2\pi \mu_q^2 }{g_s} h^2_s+\frac{1}{8} \frac{h_s^2}{n_D} \sum_{\vec{v}} \frac{1}{\beta} \sum_{n} \int \frac{d^2 p_{\bot}}{(2\pi)^2}  \int \frac{d p_{\parallel}}{2\pi} \textrm{Tr} \left[ S_{\vec{v}}(p+q)  S_{\vec{v}}(p) \right]\right\}\; ,\label{b2pt}
\end{align}
where the momentum-space propagator $S_{\vec{v}}(p)$ in a  homogeneous background can be found as
\begin{align}
S_{\vec{v}}(p) =  \frac{2 }{{\bar{p}_4}^2+{p_{\parallel}}^2+ \frac{h^2_s}{n_D}\sigma^{2}}   \left[ \frac{h_s}{\sqrt{n_D}} \left( {\begin{array}{cc}
\sigma  I & 0 \\
  0 & \sigma I \\
  \end{array} } \right)-\left( {\begin{array}{cc}
0 & \gamma_{4} \left(\bar{p}_4  +ip_{\parallel}\right) \\
 \gamma_{4} \left(\bar{p}_4 -ip_{\parallel}\right)   &  0 \\
  \end{array} } \right) \right]\;.
\end{align}
The trace over spinor and internal indices can be evaluated as follows:
\begin{align}
\textrm{Tr} \left[ S_{\vec{v}}(p+q)  S_{\vec{v}}(p) \right] 
&= - 16 n_D \left\{ \frac{1}{{\bar{p}_4}^2+E_{p+q}(\sigma)^2} + \frac{1}{{\bar{p}_4}^2+E_{p}(\sigma)^2} - \frac{1 }{[{\bar{p}_4}^2+E_{p+q}(\sigma)^2] [{\bar{p}_4}^2+E_{p}(\sigma)^2]}   \left(q_{\parallel}^2 + 4\frac{h_s^2}{n_D}\sigma^2  \right) \right\}\; ,
\end{align}
where an additional factor $1/2$ has been taken into account due to  the double-counting of spinor degrees of freedom. 
After summation over Matsubara frequencies, the bosonic two-point function~\eqref{b2pt} can be expressed as follows:
\begin{align}
\frac{\Gamma^{m (2)}_{(\delta\sigma)^2}(q_{\parallel})}{\beta (2\pi)^3 \delta^{(3)}(0)} 
&= \frac{ 2\pi \mu_q^2 }{g_s} h_s^2-\frac{\mu_q^2 h_s^2}{\pi}  \int \frac{d p_{\parallel}}{2\pi} \Bigg(\frac{1}{2E_{p}(\sigma)}\left\{ 1-n_{\textrm{f}}\left[\beta \left( E_{p}(\sigma)+\tilde{\mu}(\omega_{\parallel\mu})\right) \right]-n_{\textrm{f}}\left[\beta \left(E_{p}(\sigma)-\tilde{\mu}(\omega_{\parallel\mu})\right) \right] \right\} \nonumber\\
&\qquad\qquad\qquad\qquad \qquad~~+\frac{1}{2E_{p+q}(\sigma)}\left\{ 1-n_{\textrm{f}}\left[\beta \left( E_{p+q}(\sigma)+\tilde{\mu}(\omega_{\parallel\mu})\right) \right]-n_{\textrm{f}}\left[\beta \left(E_{p+q}(\sigma)-\tilde{\mu}(\omega_{\parallel\mu})\right) \right] \right\} \Bigg)\nonumber\\
&-\frac{\mu_q^2 h_s^2}{\pi}\left(q_{\parallel}^2 + 4\frac{h_s^2}{n_D}\sigma^2  \right)\nonumber\\
&\qquad\qquad\times  \int \frac{d p_{\parallel}}{2\pi} \frac{1}{q_{\parallel}(p_{\parallel}+q_{\parallel})}  \Bigg( -\frac{1}{2E_{p}(\sigma)}   \left\{ 1-n_{\textrm{f}}\left[\beta \left(E_{p}(\sigma)+\tilde{\mu}(\omega_{\parallel\mu})\right) \right]-n_{\textrm{f}}\left[\beta \left(E_{p}(\sigma)-\tilde{\mu}(\omega_{\parallel\mu})\right) \right] \right\} \nonumber\\
&\qquad\qquad\qquad\qquad\quad~+\frac{1}{2E_{p+q}(\sigma)}  \left\{ 1-n_{\textrm{f}}\left[\beta \left(E_{p+q}(\sigma)+\tilde{\mu}(\omega_{\parallel\mu})\right) \right]-n_{\textrm{f}}\left[\beta \left(E_{p+q}(\sigma)-\tilde{\mu}(\omega_{\parallel\mu})\right) \right] \right\} \Bigg)\;.\label{b2pt2}
\end{align}
In the limit $\mu_q \gg q_{\parallel}$, the two vacuum terms in the first integral in Eq.~\eqref{b2pt2} can be regarded as identical and regularized like in the thin-shell limit of the potential~\eqref{uvgs}:
\begin{align} 
\frac{ 2\pi \mu_q^2 }{g_s} h_s^2-\frac{\mu_q^2 h_s^2}{\pi^2}  \int_0^{\mu_q} d p_{\parallel} \frac{1}{E_{p}(\sigma)} =\frac{\mu_q^2 h_s^2}{\pi^2}  \ln \left[ \frac{\sigma_0}{\sigma} \frac{\left(1+\frac{h_s^2}{\mu_q^2}\frac{\sigma^2 }{n_D} \right)^{\frac{1}{2}}-1}{\left(1+\frac{h_s^2}{\mu_q^2}\frac{\sigma_{0}^2 }{n_D} \right)^{\frac{1}{2}}-1} \right]
\;,
\end{align}
where the constant terms with a $\sigma_{0}$-dependence are omitted in the limit $ \sigma_{0}/\mu_q \ll1 $. 
Now one can summarize the result for the two-point function as follows
\begin{align}
\frac{\Gamma^{m (2)}_{(\delta\sigma)^2}(q_{\parallel})}{\beta (2\pi)^3 \delta^{(3)}(0)} 
&\simeq \frac{\mu_q^2 h_s^2}{\pi^2} \left\{  \ln \left[ \frac{\frac{h_s}{\mu_q}\frac{\sigma_{0} }{\sqrt{n_D}} }{-1+\left(1+\frac{h_s^2}{\mu_q^2}\frac{\sigma_{0}^2 }{n_D} \right)^{\frac{1}{2}}} \right]-\ln \left[ \frac{\frac{h_s}{\mu_q}\frac{\sigma }{\sqrt{n_D}} }{-1+\left(1+\frac{h_s^2}{\mu_q^2}\frac{\sigma^2 }{n_D} \right)^{\frac{1}{2}}} \right] \right\}\nonumber\\
&\qquad\quad +\frac{\mu_q^2 h_s^2}{\pi^2}  \int_0^{\mu_q} d p_{\parallel} \frac{1}{E_{p}(\sigma)}\left\{ n_{\textrm{f}}\left[\beta \left( E_{p}(\sigma)+\tilde{\mu}(\omega_{\parallel\mu})\right) \right] + n_{\textrm{f}}\left[\beta \left(E_{p}(\sigma)-\tilde{\mu}(\omega_{\parallel\mu})\right) \right] \right\} \nonumber\\
&\qquad\quad-\frac{\mu_q^2 h_s^2}{\pi}\left(q_{\parallel}^2 + 4\frac{h_s^2}{n_D}\sigma^2  \right)\nonumber\\
&\qquad\qquad~\times  \int \frac{d p_{\parallel}}{2\pi} \frac{1}{q_{\parallel}(p_{\parallel}+q_{\parallel})} \Bigg( -\frac{1}{2E_{p}(\sigma)}   \left\{ 1-n_{\textrm{f}}\left[\beta \left(E_{p}(\sigma)+\tilde{\mu}(\omega_{\parallel\mu})\right) \right]-n_{\textrm{f}}\left[\beta \left(E_{p}(\sigma)-\tilde{\mu}(\omega_{\parallel\mu})\right) \right] \right)\}\nonumber\\
&\qquad\qquad\qquad\qquad\qquad+\frac{1}{2E_{p+q}(\sigma)}  \left\{ 1-n_{\textrm{f}}\left[\beta \left(E_{p+q}(\sigma)+\tilde{\mu}(\omega_{\parallel\mu})\right) \right]-n_{\textrm{f}}\left[\beta \left(E_{p+q}(\sigma)-\tilde{\mu}(\omega_{\parallel\mu})\right) \right] \right\} \Bigg)\;.\label{b2pt3}
\end{align}

\begin{figure}
\includegraphics[height=5.8cm]{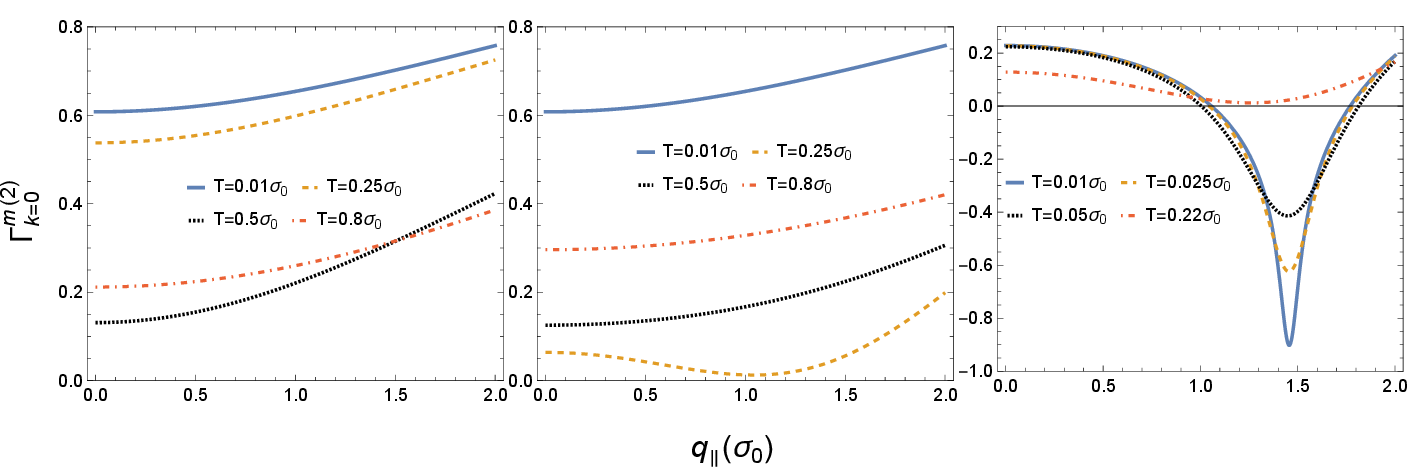} 
\caption{(Color online) $\Gamma^{m (2)}_{(\delta\sigma)^2}(q_{\parallel})$ as a function of $q_{\parallel}$ (in units of $\sigma_0$) for  $\mu=600~\textrm{MeV}$ and various temperatures: $T= 0.01\sigma_{0}\; (\textrm{blue solid}),~0.25\sigma_{0}\; (\textrm{yellow dashed}),~0.5\sigma_{0}\;(\textrm{black dotted}),~ 0.8\sigma_{0}\;(\textrm{orange dot-dashed})$. 
Left: $\bar{\mu}= 0.1 \sigma_{0}$. Middle:  $\bar{\mu}= 0.65 \sigma_{0}$. 
Right: $T= 0.01\sigma_{0}\;(\textrm{blue solid}), 0.025\sigma_{0}\; (\textrm{yellow dashed}), 0.05\sigma_{0}\;(\textrm{black dotted}), 0.22\sigma_{0}\; (\textrm{orange dot-dashed})$, for $\bar{\mu}= 0.7 \sigma_{0}$. 
The non-trivial minimum $\bar{\sigma}$ obtained with for a given ($\bar{\mu},T$) is used for each case.}\label{fig7}
\end{figure}

First, we investigate the behavior of $\Gamma^{m (2)}_{(\delta\sigma)^2} (q_{\parallel})$ in the small-$q_{\parallel}$ region. 
As one can see in Fig.~\ref{fig7}, if one investigates the regime where the amplitude of the nontrivial minimum $\bar{\sigma}$ varies smoothly ($\bar{\mu}=0.1 \sigma_{0}$, $T= 0.01\sigma_{0}, 0.25\sigma_{0}, 0.5\sigma_{0}, 0.8\sigma_{0}$),  $\Gamma^{m (2)}_{(\delta\sigma)^2} (q_{\parallel})$ shows a monotonically increasing behavior.
On the other hand, in the regime where $\bar{\sigma}$ varies strongly ($\bar{\mu}=0.65 \sigma_{0}$, $T= 0.01\sigma_{0}, 0.25\sigma_{0}, 0.5\sigma_{0}, 0.8\sigma_{0}$), $\Gamma^{m (2)}_{(\delta\sigma)^2} (q_{\parallel})$ exhibits a minimum near the phase boundary ($T=0.25\sigma_{0}$). 
Near the first-order transition regime, e.g., for $\bar{\mu}=0.7\sigma_{0}$ and $T\simeq 0.01 \sigma_{0}$, $\Gamma^{m (2)}_{(\delta\sigma)^2} (q_{\parallel})$ becomes negative in a certain range of $q_\parallel$, which  persists until $T \lesssim 0.22 \sigma_{0}$ (cf.\ right plot of Fig.~\ref{fig7}). 

\begin{figure}
\includegraphics[height=5.8cm]{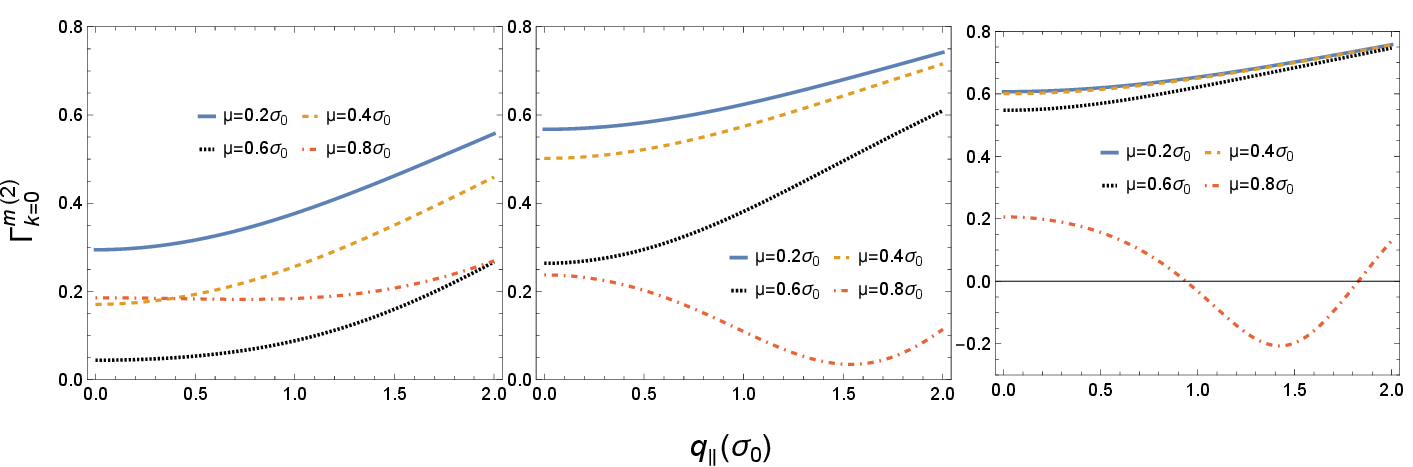} 
\caption{(Color online)   $\Gamma^{m (2)}_{(\delta\sigma)^2} (q_{\parallel})$ as a function of $q_{\parallel}$ (in units of $\sigma_0$).
We fix $\mu=600~\textrm{MeV}$ and  show various  values of $\bar{\mu}= 0.2\sigma_{0}\;(\textrm{blue solid}),~ 0.4\sigma_{0}\; (\textrm{yellow dashed}),~0.6\sigma_{0}\;(\textrm{black dotted}),~0.8\sigma_{0}\; (\textrm{orange dot-dashed})$. 
From left to right, the  plots correspond to $T= 0.2 \sigma_{0}$, $T= 0.1 \sigma_{0}$, and $T= 0.01 \sigma_{0}$, respectively. 
The value of non-trivial minimum $\bar{\sigma}$ obtained with for a given ($\bar{\mu},T$) is used for each case. }\label{fig8}
\end{figure}

To investigate the first-order transition regime, one fixes the temperature below $T=0.4\sigma_{0}$ and varies $\bar{\mu}$. 
As one anticipates from the phase diagram, $\Gamma^{m (2)}_{(\delta\sigma)^2} (q_{\parallel})$ does not show a clear instability at $T=0.2\sigma_{0}$ but it shows $\Gamma^{m (2)}_{(\delta\sigma)^2} (q_{\parallel})\simeq 0$ at $\bar{\mu}=0.8\sigma_{0}$ (cf.\ middle panel of Fig.~\ref{fig8}). 
As the temperature decreases further ($T=0.01\sigma_{0}$), $\Gamma^{m (2)}_{(\delta\sigma)^2} (q_{\parallel})$ shows an instability  just to the right of the phase boundary ($\bar{\mu}>0.7 \sigma_0)$. 

\begin{figure}
\includegraphics[height=6cm]{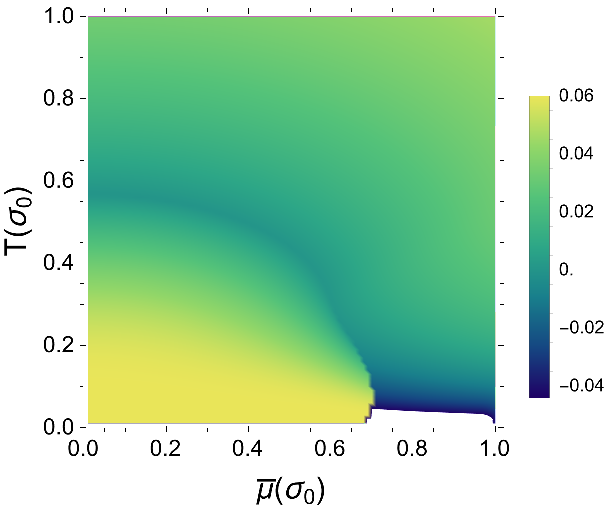} 
\includegraphics[height=6cm]{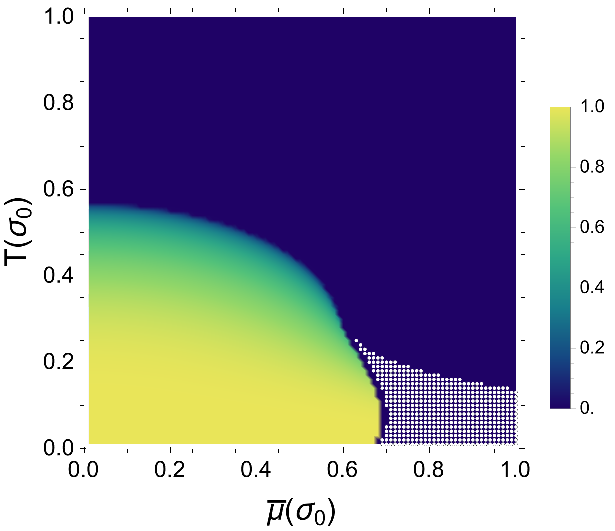}
\caption{(Color online) Left: contour plot of the minimum value of $\Gamma^{m (2)}_{(\delta\sigma)^2} (q_{\parallel})$ in the $(\bar{\mu},T)$-plane for $\mu=600~\textrm{MeV}$. 
In the white region at small $T$, $\Gamma^{m (2)}_{(\delta\sigma)^2} (q_{\parallel}) \rightarrow - \infty$.
Here, a spatial oscillation of the condensate with wave vector $ 2\bar{\mu} \vec{v}$ is expected. 
Right: contour plot of the $\bar{\sigma}$ condensate. The regime where $\Gamma^{m (2)}_{(\delta\sigma)^2} (q_{\parallel})<0$, and thus an inhomogeneous order parameter is expected, is shown as blue-dotted region. }\label{fig9}
\end{figure}

The most unstable point, which becomes singular in the limit $T\rightarrow0$, appears at $q_{\parallel} \simeq 2\bar{\mu}$. 
As the shell-thickness $\bar{\mu}$ becomes larger in the low-$T$ regime, the system undergoes a first-order phase transition and the homogeneous background field $\bar{\sigma}$ vanishes. 
The momentum $q_{\parallel}$ can be understood as the momentum difference between a quark and a hole state when the spatial fluctuation is allowed. 
A possible configuration is given in Fig.~\ref{fig6}: if the hole state is located just below the Fermi surface, $q_{\parallel}$ corresponds to the momentum carried by the excited quark. 
Once the first-order phase transition occurs in the low-$T$ regime, ($\bar{\sigma}\rightarrow 0$, as one can see from Fig.~\ref{fig9}), the wave vector $ 2\bar{\mu} \vec{v}$ is related to the appearance of another oscillating mode~\cite{Buballa:2018hux, Koenigstein:2021llr}:
\begin{align}
\langle \bar{\psi} (x) \psi (x) \rangle =  \sum_{\vec{v}} e^{2i\mu_q \vec{v} \cdot \vec{x}} [\bar{\sigma} + \delta \sigma(x)]\mathit{\Delta}_{\bot}\quad \Longrightarrow  \quad \sum_{\vec{v}} e^{2i(\mu_q +\bar{\mu}) \vec{v} \cdot \vec{x}}  \kappa \mathit{\Delta}_{\bot}=4\pi \cos(2\mu \vert \vec{x}\vert )\kappa \mathit{\Delta}_{\bot}\;,\label{ihqc}
\end{align}
if one approximates $\delta \sigma(x)$ as follows:
\begin{align}
\delta \sigma(x) = \int \frac{dq_{\parallel}}{2\pi} e^{i q_{\parallel} \vec{v}\cdot \vec{x}}\delta \sigma (q_{\parallel}) \mathit{\Delta}_{\bot}  \simeq    e^{2i\bar{\mu} \vec{v}\cdot \vec{x}} \kappa \mathit{\Delta}_{\bot}\;,
\end{align}
where $\delta \sigma (q_{\parallel}) \simeq  \delta(q_{\parallel}-2\bar{\mu})\kappa$.
The regime where a homogeneous $\bar{\sigma}$ condensate becomes unstable is shown in Fig.~\ref{fig9}.
This phase diagram looks rather similar to the one of the standard GN model \cite{Schnetz:2004vr,Koenigstein:2021llr}, except that the ordinary chemical potential $\mu_q$ is replaced by the shell-thickness parameter $\bar{\mu}$.

\subsection{Incorporating $\sigma$ fluctuations via FRG}

In the previous subsections, we only considered the bosonic degrees of freedom as a mean  background field or the boson mode generated by quantum fluctuations of the quark Fermi surface. 
Now we investigate the case where quantum fluctuations of the $\sigma$-mode are accounted for.
Thus, we add the following kinetic term to the effective average action,
\begin{align}
\Gamma^{b} = \frac{1}{2}  \sum_{\vec{v}} \int d^4x_{\textrm{E}} \partial_{\mu} \sigma \partial_{\mu} \sigma\;.\label{efeq2}
\end{align}
The effective average action underlying the RG-time running of the effective potential $U_{\sigma}(k, \sigma)$ reads as follows:
\begin{align}
\Gamma_{k} =\int d^4x_{\textrm{E}} \bigg\{ & \frac{1}{2}\sum_{\vec{v}}\bigg[\bar{\Psi}_{\vec{v}}\left (-i  \Gamma_{\mu} \partial_{\mu} -\bar{\mu}\Gamma^0+ \frac{h_s}{\sqrt{n_D}}\varphi + \frac{h_v}{\sqrt{n_D}}\omega_{\mu} \Gamma_{\mu} \right) \Psi_{\vec{v}} +\partial_{\mu} \sigma \partial_{\mu} \sigma\bigg] + 2 \pi \frac{\mu_q^2 h^2_v}{g_v}(\omega_4^2-\omega_{\parallel}^2)+ U_{\sigma}(k, \sigma)\bigg\} \nonumber\\
& + \Delta \Gamma^{f}_{k} + \Delta \Gamma^{b}_{k}\;,\label{Ab2}
\end{align}
where $U_{\sigma}(k, \sigma)$ is assumed to depend only on the amplitude of $\sigma^{+}$ field ($\sigma = \sqrt{(\sigma^{+})^{\dagger}\sigma^{+}}$). The bosonic regulator term $\Delta \Gamma^{b}_{k}$ is given as
\begin{align}
\Delta \Gamma^{b}_{k} = \frac{1}{2}\sum_{\vec{v}}  \int \frac{d^4p_{\textrm{E}}}{(2\pi)^4} \sigma(-p_{\parallel}) R^{b}_{\vec{v}}(k,p) \sigma(p_{\parallel})\;. 
\end{align}
This approximation corresponds to the general form of the local potential approximation (LPA) where the RG running of the kinetic terms are ignored~\cite{Morris:1994ki, Hasenfratz:1985dm, Zumbach:1994kc}. 
Since the LPA can be understood as keeping the lowest-order contribution of the derivative expansion, this truncation scheme is suitable for systems where  small momentum-exchange interactions  are dominant. 
Here we assume that the bosonic fluctuation is absent in the UV limit ($\Gamma_{k} = \Gamma^{m}$ for $k\rightarrow \mu_q$) and is generated during the RG running towards the IR limit. 
Thus, one can impose the following UV initial condition $U_{\sigma}(k\rightarrow \mu_q, \sigma)$: 
\begin{align}
U_{\sigma}(\mu_q, \sigma)&= \frac{2\pi\mu_q^2 (h_s \sigma)^2}{g_s} =
-\frac{\mu_q^2 (h_s \sigma)^2}{\pi^2} \left\{  \left(1+\frac{h_s^2}{n_D}\frac{\sigma_{0}^2 }{\mu_q^2} \right)^{-\frac{1}{2}}-\ln \left[ \frac{\frac{h_s}{\sqrt{n_D}}\frac{\sigma_{0} }{\mu_q} }{\left(1+\frac{h_s^2}{n_D}\frac{\sigma_{0}^2 }{\mu_q^2} \right)^{\frac{1}{2}}-1} \right] \right\}\;,
\end{align}
where the UV coupling~\eqref{uvgs} from the mean-field approximation has been used. 

The exact RG flow equation including the contribution from the dynamical $\sigma$ mode reads as follows: 
\begin{align}
\partial_t \Gamma_k =- \textrm{STr} \left\{ \frac{\partial}{\partial t} R^{f}_{\vec{v}}(k,p) \left[ 
\Gamma^{m (2)}_{\bar{\Psi}\Psi}(k, \vec{v})+R^{f}_{\vec{v}}(k,p) \right]^{-1} \right\} + \textrm{STr} \left\{ \frac{\partial}{\partial t} R^{b}_{\vec{v}}(k,p) \left[ \Gamma^{b(2)}_{\sigma^2}(k,\vec{v})+R^{b}_{\vec{v}}(k,p) \right]^{-1} \right\}\;,\label{efb}
\end{align}
where $\Gamma^{b(2)}_{\sigma^2}(k,\vec{v})$  denotes the second-order functional derivative of $\Gamma_k$ with respect to $\sigma$. 
The bosonic Litim regulator $R^{b}_{\vec{v}}(k,p)$ is 
\begin{align}
R^{b}_{\vec{v}}(k,p)= \frac{1}{2}\beta\delta_{n'n} \delta^{(2)}(\vec{v}^{\, \prime}-\vec{v})\, 2\pi \delta(p_{\parallel}' - p_{\parallel}) p^2_{\parallel}r_b(k,p_{\parallel})\;,
\end{align}
which leads to
\begin{align}
 \left[ \Gamma^{b(2)}_{\sigma^2}(k,\vec{v})+R^{b}_{\vec{v}}(k,p) \right]^{-1}=\frac{ 2 \beta\delta_{n'n} \delta^{(2)}(\vec{v}^{\, \prime}-\vec{v}) \,2\pi \delta(p_{\parallel}' - p_{\parallel}) }{ p_4^2 + p_{\parallel}^2[ 1+r_b(k,p_{\parallel})]+\frac{\mu_q^2 h^2_s}{g_s}}\;.
\end{align}
The $\sigma$-part of the exact RG flow equation can be obtained as follows:
\begin{align}
\textrm{STr} \left\{ \frac{\partial}{\partial t} R^{b}_{\vec{v}}(k,p) \left[ \Gamma^{b(2)}_{\sigma^2}(k,\vec{v})+R^{b}_{\vec{v}}(k,p) \right]^{-1} \right\}=- \beta (2\pi)^3 \delta^{(3)}(0) \frac{ \mu_q^2}{\pi^2} \frac{ k(t)^3  }{ 2 E^b_k(\partial^2_{\sigma}U_{\sigma}) }\left\{ 1+ \frac{2}{\exp\left[\beta\left(E^b_k(\partial^2_{\sigma}U_{\sigma}) \right)\right]-1} \right\}\;,
\end{align}
where $E^b_k(\partial^2_{\sigma}U_{\sigma}) = \left[  k(t) ^2+ \partial^2_{\sigma}U_{\sigma} \right]^{\frac{1}{2}} $. Equation~\eqref{efeq2} can then be written as
\begin{align}
\partial_t \Gamma_k &= \beta (2\pi)^3\delta^{(3)}(0)  \frac{2n_D \mu_q^2}{\pi^2} \frac{ k(t)^3  }{E_k(\sigma)}\left\{1-n_{\textrm{f}}\left[\beta \left(E_k(\sigma)+\tilde{\mu}(\omega_{\parallel\mu})\right) \right] - n_{\textrm{f}}\left[\beta \left(E_k(\sigma)-\tilde{\mu}(\omega_{\parallel\mu})\right) \right] \right\}\nonumber\\
&\qquad - \beta (2\pi)^3 \delta^{(3)}(0) \frac{ \mu_q^2}{\pi^2} \frac{ k(t)^3  }{ E^b_k(\partial^2_{\sigma}U_{\sigma})}\left\{1+2n_{\textrm{b}}\left[\beta\left(2E^b_k(\partial^2_{\sigma}U_{\sigma}) \right) \right]  \right\}\;,\label{flow}
\end{align} 
where $n_{\textrm{b}}[x]=1/(e^x-1)$.

Considering that the only scale-dependent quantity in the ansatz~\eqref{Ab2} is the effective potential $U_\sigma(k, \sigma)$, plugging \cref{Ab2} into the left-hand side of \cref{flow} we  obtain 
\begin{align}
\partial_t U_\sigma(k, \sigma) &=  \frac{2n_D \mu_q^2}{\pi^2} \frac{ k(t)^3  }{E_k(\sigma)}\left\{1-n_{\textrm{f}}\left[\beta \left(E_k(\sigma)+\tilde{\mu}(\omega_{\parallel\mu})\right) \right] - n_{\textrm{f}}\left[\beta \left(E_k(\sigma)-\tilde{\mu}(\omega_{\parallel\mu})\right) \right] \right\}\nonumber\\
&\qquad -  \frac{ \mu_q^2}{\pi^2} \frac{ k(t)^3  }{ E^b_k(\partial^2_{\sigma}U_{\sigma})}\left\{1+2n_{\textrm{b}}\left[\beta\left(2E^b_k(\partial^2_{\sigma}U_{\sigma}) \right) \right]  \right\}\;.\label{flow2}
\end{align} 
This flow equation closely resembles the one which is derived for the GN model in the (1+1)-dimensional case~\cite{Stoll:2021ori}. 
In particular, we can follow this analogy and use a similar approach to the one of Ref.~\cite{Stoll:2021ori} for the solution of the flow equation. 

We first observe that the RG-time evolution of the effective potential, given by the right-hand side of Eq.~\eqref{flow2}, is independent of the effective potential itself, while it depends on the second derivative of the effective potential. 
This suggests to use its derivative with respect to the $\sigma$-field as a new variable. 
Thus, we  define the new variable
\begin{align}
   u(k,\sigma)\equiv \partial_\sigma U_\sigma(k,\sigma)\;,
\end{align}
 and analogously
\begin{align}
   u'(k,\sigma)\equiv \partial_\sigma u(k,\sigma)\equiv \partial^2_{\sigma}U_\sigma(k,\sigma)\;.
\end{align}
In order to express Eq.~\eqref{flow2} in terms of the newly defined variables, we now take the derivative with respect to $\sigma$ of Eq.~\eqref{flow2} and obtain
\begin{align}
\partial_t u(k, \sigma) &=  \frac{d}{d\sigma}\left(\frac{2n_D \mu_q^2}{\pi^2} \frac{ k(t)^3  }{E_k(\sigma)}\left\{1-n_{\textrm{f}}\left[\beta \left(E_k(\sigma)+\tilde{\mu}(\omega_{\parallel\mu})\right) \right] - n_{\textrm{f}}\left[\beta \left(E_k(\sigma)-\tilde{\mu}(\omega_{\parallel\mu})\right) \right] \right\}\right)\nonumber\\
&\qquad - \frac{d}{d\sigma}\left( \frac{ \mu_q^2}{\pi^2} \frac{ k(t)^3  }{ E^b_k(u'(k,\sigma))}\left\{1+2n_{\textrm{b}}\left[\beta\left(2E^b_k(u'(k,\sigma)) \right) \right]  \right\}\right)\;,\label{flow3}
\end{align} 
which we can rewrite as
\begin{align}
\partial_t u(k,\sigma) + \frac{d}{d\sigma} g(t, u'(k,\sigma)) =\frac{d}{d\sigma} S(t, \sigma)\;.\label{flow_sigma_a}
\end{align}

Equation~(\ref{flow_sigma_a}) is a non-linear diffusion equation, like the heat equation, with an additional source term induced by the fermions.  
If we let us then guide by the  interpretation of the FRG flow equation as a diffusion equation, $t$ can be considered as a time variable for the FRG flow, while $\sigma$ can be interpreted as a spatial variable. 
In this specific case $u(k, \sigma)$ plays the role of a fluid ``density", whose evolution properties are governed by the two contributions we introduced in \cref{flow_sigma_a} and that we are about to detail.
It is clear that, once the association between the FRG flow equation and a diffusion equation has been demonstrated, we are then allowed to exploit the wide and well-established toolbox of powerful numerical methods that have been developed to solve hydrodynamic equations.  
In particular, we will use the so-called Kurganov and Tadmor (KT) scheme~\cite{Kurganov:2000ovy}.

First of all, the radial $\sigma$-mode produces the  diffusion term
\begin{align}
	g(t,  u')=\frac{ \mu_q^2}{\pi^2} \frac{ k(t)^3  }{ E^b_k(u'(k,\sigma))}\left\{1+2n_{\textrm{b}}\left[\beta\left(2E^b_k(u'(k,\sigma) \right) \right] \right\}\;,
\end{align}
since it depends on the curvature mass $u'(k,\sigma)$. 
From a fluid-dynamical perspective, diffusion leads to a smearing of the solution, since it depends on the gradient of the solution itself.
This implies that the conserved quantity is transported from regions in the spatial domain where it is more concentrated, i.e., where the  solution significantly differs from a constant one, to regions where it is less concentrated, i.e., where it is closer to a constant one, which by definition is left unchanged by diffusion.
According to these features, the diffusion term plays a fundamental role in the dynamics of symmetry restoration.  

On the other hand, the fermionic loop gives rise to a time- and $\sigma$-dependent source term
\begin{align}\label{fermions}
    S(t,\sigma) =\frac{2n_D \mu_q^2}{\pi^2} \frac{ k(t)^3  }{E_k(\sigma)}\left\{1-n_{\textrm{f}}\left[\beta \left(E_k(\sigma)+\tilde{\mu}(\omega_{\parallel\mu})\right) \right] - n_{\textrm{f}}\left[\beta \left(E_k(\sigma)-\tilde{\mu}(\omega_{\parallel\mu})\right) \right] \right\}\;,
\end{align} 
which is independent of the conserved quantity $u(k,\sigma)$. 
According to this observation, the fermionic contribution turns out to be completely independent of the effective potential and of its time evolution, so it receives no feedback from the bosonic sector.  
 
If one considers the case $\tilde{\mu}=0$, and performs the $\sigma$-derivative of Eq.~\eqref{fermions}, as indicated in Eq.~\eqref{flow_sigma_a},  one finds a source-like, positive contribution to the flow equation for $\sigma < 0 $ and a sink-like negative contribution for $\sigma> 0$.  
This is not trivially true for $\tilde{\mu}\neq0$, and also some high peaks and shocks in the field space may develop, especially in the low-temperature case~\cite{Stoll:2021ori}.

\begin{figure}
\includegraphics[height=6cm]{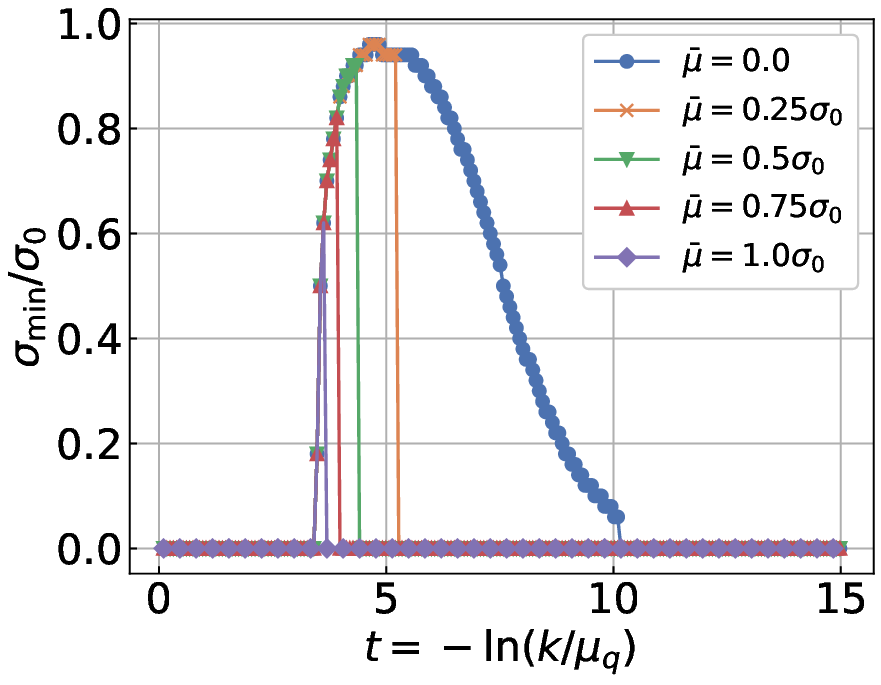}
\includegraphics[height=6cm]{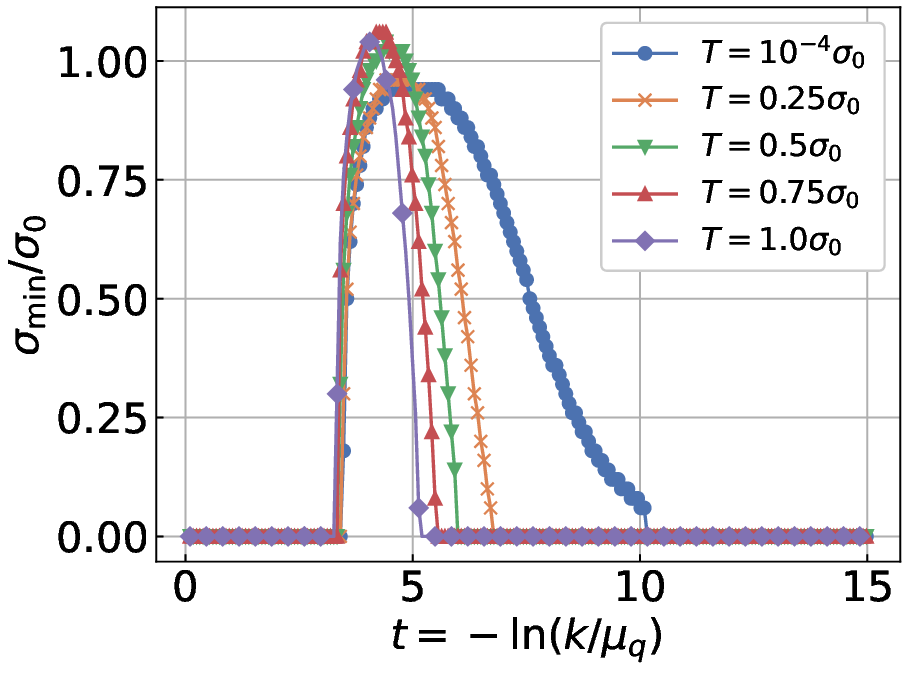}
\caption{(Color online) Scaled condensate (position of the minimum)
$\sigma_{\mathrm{min}} (t)/\sigma_0$  as a function of  $t=-\ln(k/\mu_q)$ with $\mu_q=40\sigma_0$. 
Left: at a given constant temperature $T = 10^{-4}\sigma_0$ but for different values of shell thickness $\bar{\mu}$. 
Right: in the thin-shell limit  $\bar{\mu}= 0.0$, but for different values of temperature $T$.  Other relevant parameters  are: $n_D=6$, $h_s=h_v\equiv\sqrt{n_D}$, $\omega_0=0.0$, $\omega_\parallel=0.2\bar{\mu}/\pi^2$, and $\sigma_0=10$ MeV, respectively.}\label{fig11}
\end{figure}

The results obtained within the FRG approach show that the effective potential develops a  nonzero minimum at a finite RG scale, which breaks the $\mathbf{Z}_2$ symmetry of the potential. 
However, at a later RG scale the symmetry is restored, i.e., the $\sigma$-condensate vanishes. 
The symmetry restoration is independent on both the shell thickness and the temperature, and increasing these parameters results in a higher RG scale at which the symmetry is restored.
From a fluid-dynamical point of view, this occurrence can be interpreted as follows:
in the early stages of the RG-time evolution, a finite condensate is produced by the action of the sink-like fermionic term. 
At later RG times, however, the action of the  bosonic contribution becomes dominant, implying that the diffusion term smears out the potential up to the point where the symmetry is restored.  
This result is in agreement with the previous findings in the context of the GN model in (1+1) dimensions~\cite{Stoll:2021ori}. 
This result means that, if one considers the bosonic $\sigma$-mode as dynamical degree of freedom, restoration of the $\mathbf{Z}_2$ symmetry is inevitable and the nontrivial shell-like structure where $\bar{\sigma}\neq0$ is not present in the IR limit, in contrast to the mean-field analysis of the simple bosonized model~\ref{Ab}. 
However, we cannot discard the possibility that higher-order truncations beyond LPA, or including fluctuations of the vector meson $\omega_\mu$ in LPA, may lead to different results.
So far, our conclusion is thus that the inclusion of bosonic fluctuations leads to an inevitable restoration of the $\mathbf{Z}_2$ symmetry, implying that, in contrast to the  mean-field results, no finite condensate is present in the IR.

\section{Conclusion}~\label{sec4}

In this work, we constructed a simple effective model where the collinear four-quark interactions in high-density effective theory (HDET) are further reduced to contact interactions analogous to the GN model. 
The contact interaction can be related to QCD via the collinear four-quark interaction around the Fermi surface in an instanton background. 
By considering the fluctuation in the direction transverse to a given Fermi velocity, transitions between two topological different vacua ($I=\pm1$) are possible via the surface quark and hole mode on opposite sides of the Fermi sphere. 
The corresponding scalar channel can form an Overhauser-type surface quark-hole condensate $\sigma$, which corresponds to the amplitude of the inhomogeneous scalar quark condensation:
\begin{align}
\langle \bar{\psi} (x) \psi (x) \rangle =  \sum_{\vec{v}} e^{2i\mu_q \vec{v} \cdot \vec{x}} \langle \bar{\psi}_{+} (-\vec{v}, x) \psi_{+} (\vec{v}, x) \rangle \quad \Longrightarrow \quad 4 \pi \cos(2\mu_q \vert \vec{x} \vert) \sigma \mathit{\Delta}_{\bot}\;.
\end{align}
The vector channel was included to check the role of a repulsive interaction in the scalar-condensation process. 
In the construction of the effective model, the relevant slow modes are defined with respect to the ground level defined at $\mu_{q}=\mu - \bar{\mu}$, and the condensation pattern is analyzed as a function of the shell thickness parameter $\bar{\mu}$ and the temperature $T$. 
The non-perturbative RG flow of the effective average action is first analyzed in the mean-field approximation, i.e., where the RG flow equation can be integrated analytically.

In the background mean-field approximation for the collective bosonic fields, the phase diagram for a homogeneous condensate is found in the ($\bar{\mu}$, $T$)-plane (see Fig.~\ref{fig5}). 
To the left side of a tricritical point at ($\bar{\mu} = 0.63 \sigma_{0}$, $T = 0.25 \sigma_{0}$), a continuous variation of the order parameter is found, which goes to zero in a second-order phase transition. 
To the right of this point, the transition is of first order.
At larger values of $\bar{\mu}$, beyond the phase boundary for the first-order transition, one can find a regime where an inhomogeneous configuration of $\sigma(x)$ is energetically favored over the configuration with a homogeneous condensate.
This conclusion follows by analyzing the second-order coefficient of the effective average action, which becomes negative (Fig.~\ref{fig9}). 
If one assumes the fluctuation $\delta \sigma(x)$ to have the main contribution from the wave vector $2\bar{\mu}\vec{v}$, as one may expect from Fig.~\ref{fig7}, the inhomogeneous scalar quark condensation varies spatially with a frequency $2\mu \equiv 2 (\mu_q + \bar{\mu})$ (Eq.~\eqref{ihqc}),
\begin{align}
\langle \bar{\psi} (x) \psi (x) \rangle \quad \Longrightarrow  \quad \sum_{\vec{v}} e^{2i(\mu_q +\bar{\mu}) \vec{v} \cdot \vec{x}}  \kappa \mathit{\Delta}_{\bot}=4\pi \cos(2\mu \vert \vec{x}\vert )\kappa \mathit{\Delta}_{\bot}\;.
\end{align}
In other words, if one assumes a temperature $T$ corresponding to a state which, in the mean-field approximation, corresponds to the homogeneously broken phase, and one then increases $\bar{\mu}$, the homogeneous condensation amplitude $\bar{\sigma}$ persists until the momentum shell reaches a thickness of about $\bar{\mu}\simeq 0.7 \sigma_{0}$.
Above this value, as the condensation amplitude has a dominant oscillation frequency $2\bar{\mu}$, the quark condensation becomes periodic with the main frequency $2\mu$. 
This configuration looks like  quarkyonic matter~\cite{McLerran:2007qj}, where the explicit baryon shell structure is deduced from large-$N_c$ QCD.

However, if one regards the collective bosonic mode $\sigma$ as a quasi-particle field by assuming the dynamical $\sigma(x)$ mode, the aforementioned nontrivial minimum $\sigma$ does not appear. 
The bosonic and fermionic part of the FRG flow equation correspond to a diffusion and source term, respectively. 
In the FRG flow, a nontrivial condensation amplitude is formed by the fermionic sink-like term but, as the bosonic diffusion term becomes dominant in the later RG evolution, the condensate vanishes eventually. 
While it is then questionable whether an inhomogeneous configuration exists at all, the possibility that it exists is not ruled out and needs to be explored in future work.

Other extensions of the current work consist of extending the FRG analysis to the LPA' approximation, where a non-trivial wavefunction renormalization for the fields is considered~\cite{Dupuis:2020fhh}, or by including fluctuations of the vector mesons in LPA and beyond.
This may potentially change the conclusions of the present work.

\begin{acknowledgements}
The authors gratefully acknowledge fruitful discussions with L.\ Kiefer, A.\ Koenigstein, and R.D.\ Pisarski.
K.S.\ acknowledges support of the Alexander von Humboldt foundation.
The authors acknowledge support by the Deutsche Forschungsgemeinschaft (DFG, German Research Foundation) through the CRC-TR 211 ''Strong-interaction matter under extreme conditions'' -- project number 315477589 -- TRR 211.
The work is supported by the State of Hesse within the Research Cluster ELEMENTS (Project ID 500/10.006). 

\end{acknowledgements}

\appendix

\section{Euclidean convention}

In this work, we follow the notations given in Ref.~\cite{Bellac:2011kqa}. From Minkowski space to Euclidean space, the time components are transformed as follows: $t\rightarrow -i\tau$, $\gamma_0\rightarrow -i \gamma_4$, $p_0\rightarrow -ip_4 = i\omega_n$, where $\omega_n$ denotes Matsubara frequencies. 
Note following subsequent transformations:
\begin{align}
g_{\mu\nu} \rightarrow -\delta_{\mu\nu},~ A_{\mu}B^{\mu} \rightarrow -A_{\mu}B_{\mu},~ \slash \hspace{-0.2cm}  p  \rightarrow -\slash \hspace{-0.2cm}  p_{E} = -p_4\gamma_4 - \vec{\gamma}\cdot \vec{p},~\partial^{\mu}A_{\mu}=\partial_{\mu}A_{\mu}\;.
\end{align}

 The partition function is defined as follows:
\begin{align}
\mathcal{Z}&\equiv \textrm{const.}\int \mathcal{D}[\bar{\psi}, \psi, \cdots] \exp\left[ -S_{\textrm{E}} \right],\quad S_{\textrm{E}} = \int d^4 x_{\textrm{E}} \mathcal{L}_{\textrm{E}}(\bar{\psi}, \psi, \cdots)\;.
\end{align}

\section{Instanton solution}
\label{app:B}

The Euclidean $\textrm{SU(2)}$ Yang-Mills action can be written as follows:
\begin{align} S^{\textrm{E}}_{\textrm{YM}}=\frac{1}{4}\int d^4 x_{\textrm{E}}\mathcal{G}^{a}_{\mu\nu}\mathcal{G}^{a}_{\mu\nu}=\frac{1}{4}\int d^4 x_{\textrm{E}} \left[ \pm \mathcal{G}^{a}_{\mu\nu} \tilde{\mathcal{G}}^{a}_{\mu\nu}+\frac{1}{2}\left( \mathcal{G}^{a}_{\mu\nu}\mp  \tilde{\mathcal{G}}^{a}_{\mu\nu}\right)^2 \right]\;,
\end{align}
where $\pm$ denotes the instanton and anti-instanton field strength, respectively and $\tilde{\mathcal{G}}^{a}_{\mu\nu} =\frac{1}{2} \epsilon_{\mu\nu\rho\sigma}\mathcal{G}^{a}_{\rho\sigma}$ with $\epsilon_{1234}=i$. 
This action can be minimized if the field-strength tensor is (anti-)self-dual: $\mathcal{G}^{a}_{\mu\nu} =\pm \tilde{\mathcal{G}}^{a}_{\mu\nu}$ (where $\pm$ denotes self-duality and anti-self-duality, respectively). 
Under the (anti-)self-dual field-strength configuration, the minimized action $ S^{\textrm{E}}_{\textrm{YM}}=8\pi^2 \vert Q \vert/g^2 $ is determined by the topological invariant $Q$:
\begin{align}
Q&=\frac{g^2}{32\pi^2}\int d^4 x_{\textrm{E}}  \mathcal{G}^{a}_{\mu\nu} \tilde{\mathcal{G}}^{a}_{\mu\nu}=\int d^4 x_{\textrm{E}} \partial_{\mu} K_{\mu}\;,\\
K_{\mu}&=\frac{1}{16\pi^2}\epsilon_{\alpha \beta \gamma \delta}\left( A^{a}_{\beta}\partial_{\gamma} A^{a}_{\delta}+\frac{1}{3}\epsilon^{abc} A^{a}_{\beta} A^{b}_{\gamma} A^{c}_{\delta}\right)\;.
\end{align}
 To have a finite action, the gauge field should be pure gauge at the surface of a large sphere in 4D Euclidean space: $A_{\mu}\rightarrow i U \partial_{\mu} U$ as $\vert x_{\mu} \vert \rightarrow \infty$. 
 The gauge transformation $U$ links the isospin to the spatial 3-sphere and can be classified by the winding number $n_{\textrm{CS}}$. 
 If the gauge field vanishes rapidly in the large-sphere limit, the topological invariant can be understood as
\begin{align}
Q&=\int d^4 x_{\textrm{E}} \partial_{\mu} K_{\mu}= \int d^4 x_{\textrm{E}} \partial_{4} K_{4}=i \int dt \partial_{0} \int d^3 x K_{0} = n_{\textrm{CS}} (t=\infty)-n_{\textrm{CS}} (t=-\infty)\;,\label{index1}
\end{align}
where one can find the transition between topologically different vacua ($Q\neq0$) by the instanton solution. 
By taking the large-sphere limit of  $A_{\mu}\rightarrow i U \partial_{\mu} U$ with $U=i (x_{\mu}/\vert x_{\mu} \vert) \tau^{+}_{\mu}$, where $\tau^{+}_{\mu}=(\mp i, \vec{\tau})$, one can obtain the asymptotic form $A^{a}_{\mu}=2 \eta_{a\mu\nu} x_{\nu}/x^2$ for the $Q=1$ configuration. 
The group structure $\eta_{a\mu\nu}$ known as `t Hooft symbol~\cite{tHooft:1976rip} can be found as
\begin{align}
\eta_{a\mu\nu} =&
  \Bigg\{ {\begin{array}{cc}
 \epsilon_{a\mu\nu},& \mu,\nu =1,2,3\\
  \delta_{a\mu},&\nu=4,\\
   -\delta_{a\nu},& \mu=4,
  \end{array}},\quad 
  \bar{\eta}_{a\mu\nu} =
  \Bigg\{ {\begin{array}{cc}
 \epsilon_{a\mu\nu},& \mu,\nu =1,2,3\\
  -\delta_{a\mu},&\nu=4,\\
   \delta_{a\nu},& \mu=4,
  \end{array}},
\end{align}
where $\bar{\eta}_{a\mu\nu}$ stands for the group structure of the anti-instanton solution ($Q=-1$), which is embedded in another $\textrm{SO}(3)$ subgroup of $\textrm{O}(4)$. 
One can find a general solution of the self-duality relation by using the ansatz $A^{a}_{\mu}=2 \eta_{a\mu\nu} x_{\nu}f(x^2)/x^2$ with the boundary constraint $f(x^2)\rightarrow 1$ as $\vert x_{\mu} \vert \rightarrow \infty$. The solution known as Belavin-Polyakov-Schwartz-Tyupkin instanton~\cite{Belavin:1975fg} can be found as follows:
\begin{align}
A^{a}_{\mu}=\frac{2 \eta_{a\mu\nu}x_{\nu} }{x^2+\rho^2}\;,
\end{align}
where $f(x^2)=x^2/(x^2+\rho^2)$ and $\rho$ is the scale parameter for the instanton size. 
A comprehensive explanation can be found in Refs.~\cite{Schafer:1996wv, Coleman:1985rnk}.

\section{Zero-mode contributions in vacuum}\label{appc}

The physical meaning of the topological charge~\eqref{index1} can be understood from the axial anomaly relation $\partial_{\mu}j^{5}_{\mu}=(N_fg^2/16\pi^2)\mathcal{G}^{a}_{\mu\nu} \tilde{\mathcal{G}}^{a}_{\mu\nu}$. The space-time integration of the divergence of the axial current becomes
\begin{align}
\int d^4x \partial_{\mu}j^{5}_{\mu} &= \Delta Q_{5}= Q_5(t=\infty) -  Q_5(t=-\infty)= - i N_f \int d^4x_{\textrm{E}} \partial_{\mu} \textrm{Tr}\left[ S(x,x)\gamma_{5} \gamma_{\mu}\right]\;,
\end{align}
where the quark propagator $S_{\textrm{E}}(x,y) = \left\langle x \vert ( i \slash \hspace{-0.25cm} D(\mu \rightarrow 0 ))^{-1}  \vert y \right\rangle$ can be decomposed into eigenmodes as follows:
\begin{align}
S_{\textrm{E}}(x,y) = - \sum_{\lambda} \frac{\psi_{\lambda}(x) \psi^{\dagger}_{\lambda}(y)}{\lambda}\;,\label{dopeg}
\end{align}
where each eigenmode is normalized and $\lambda$ is the pure imaginary eigenvalue defined from $ i \slash \hspace{-0.25cm} D(\mu \rightarrow 0 ) \psi_{\lambda} = \lambda  \psi_{\lambda} $ (Eq.~\eqref{diracoprels}). 
By using the anti-hermiticity of $i \slash \hspace{-0.25cm} D(\mu \rightarrow 0 )$, $\Delta Q_{5}$ can be evaluated as
\begin{align}
\Delta Q_{5}&=  i N_f \int d^4x_{\textrm{E}}  \textrm{Tr}\left[ \sum_{\lambda} \frac{\psi_{\lambda}(x) \psi^{\dagger}_{\lambda}(x)}{\lambda} 2\lambda \gamma_{5} \right] = 2N_f (n_R-n_L)\;,
\end{align}
where only the zero-modes $\psi_{(L/R)0}$ contribute to the summation since  $\psi_{\lambda}$ and $\gamma_5 \psi_{\lambda} = \psi_{-\lambda}$ are orthogonal due to the anti-hermiticity of the Dirac operator. 
This relation is known as index theorem~\cite{Atiyah:1970ws, Callias:1977kg}. 
The proper version of the index theorem at $\mu \neq 0$ (Eq.~\eqref{index2}) is derived in Ref.~\cite{Kanazawa:2011tt}, and it is shown that the following relations are satisfied in most of the gauge configurations:
\begin{align}
I&=n_R(\mu)-n_L(\mu)=n_R(-\mu)-n_L(-\mu) =n_R(\mu)-n_L(-\mu)=n_R(-\mu)-n_L(\mu)\;.
\end{align}

One can find the zero-mode contribution to the quark condensate in the vacuum limit from the decomposition~\eqref{dopeg}:
\begin{align}
\left\langle \bar{\psi} \psi \right\rangle = - \textrm{Str}[S_{\textrm{E}}(x,x)] =  \int d^4x_{\textrm{E}}  \sum_{\lambda} \frac{\psi_{\lambda}(x) \psi^{\dagger}_{\lambda}(x)}{\lambda+m} =\sum_{\lambda} \frac{1}{\lambda+m}=-\sum_{\lambda \ge 0}\frac{2m}{-\lambda^2+m^2}\;,
\end{align}
where $m$ denotes the small mass which will be eventually taken to zero, $m \rightarrow 0$. 
If one takes the continuous (thermodynamic) limit first, then the eigenvalue summation can be written as continuous integration with the spectral density $\rho(\bar{\lambda})=\sum_{\lambda} \delta(\bar{\lambda}-\lambda)$:
\begin{align}
\left\langle \bar{\psi} \psi \right\rangle = - \int_{-\infty}^{\infty} d \bar{\lambda} \rho(\bar{\lambda}) \frac{m}{-\bar{\lambda}^2+m^2}=-\pi \rho (\bar{\lambda}=0)\;,\label{bc}
\end{align}
where $\pi \delta(x) = \lim_{\epsilon \rightarrow0 }  \epsilon/(x^2+\epsilon^2)$ is used. One can see that only the zero-modes are correlated in the quark condensate. 
Equation~\eqref{bc} is known as Banks-Casher relation~\cite{Banks:1979yr}.

\bibliographystyle{aipauth4-1}

\end{document}